\documentclass[letter]{aa}

\usepackage{graphicx}
\usepackage{txfonts}

\usepackage{natbib}         

\usepackage{float}

\usepackage{capt-of}

\usepackage{hyperref}
\hypersetup{colorlinks,linkcolor={blue},citecolor={blue},urlcolor={blue}} 
\makeatletter
\renewcommand*\aa@pageof{, page \thepage{} of \pageref*{LastPage}}
\makeatother

\begin{document}

\title{
First GMVA observations with the upgraded NOEMA facility: VLBI imaging of BL Lacertae in a flaring state\thanks{Input files for Appendix~\ref{sec:calim} are only available in electronic form at the CDS via anonymous ftp to {\tt cdsarc.u-strasbg.fr} (130.79.128.5) or via {\tt http://cdsweb.u-strasbg.fr/cgi-bin/qcat?J/A+A/}}
}

\titlerunning{
Imaging of BL\,Lac in first GMVA observations with the upgraded NOEMA facility
}

\author{Dae-Won Kim\inst{1},
Michael Janssen\inst{1,2},
Thomas P.\ Krichbaum\inst{1},
Bia Boccardi\inst{1},
Nicholas R. MacDonald\inst{1,3},
Eduardo Ros\inst{1},
Andrei P.\ Lobanov\inst{1}, and
J.\ Anton Zensus\inst{1}
}

\authorrunning{D. -W. Kim et al.}

\institute{Max-Planck-Institut f\"{u}r Radioastronomie, Auf dem H\"{u}gel 69, 53121 Bonn, Germany\\ \email{dwkim@mpifr-bonn.mpg.de}
%
\and
Department of Astrophysics, Institute for Mathematics, Astrophysics and Particle Physics (IMAPP), Radboud University, P.O. Box 9010, 6500 GL Nijmegen, The Netherlands
\and
Department of Physics and Astronomy, 108 Lewis Hall, University of Mississippi, Oxford, MS 38677-1848, USA
}

\date{\today}


\abstract
{
We analyze a single-epoch Global mm-VLBI Array (GMVA) observation of the blazar BL\,Lacertae (BL\,Lac) at 86\,GHz from April 2021. 
The participation of the upgraded, phased Northern Extended Millimetre Array (NOEMA) adds additional sensitivity to the GMVA, which has facilitated the imaging of BL\,Lac during an unprecedentedly strong $\gamma$-ray flare. 
We aim to explore the nature of the inner subparsec jet of BL\,Lac and the impact of the NOEMA participation in the observation. 
For the data reduction, we employed two advanced automatic pipelines: 
\texttt{rPICARD} for the flux density calibration as well as the model-agnostic signal stabilization and \texttt{GPCAL} for the antenna leakage calibration.
The conventional hybrid imaging (CLEAN\,$+$\,amplitude and phase self-calibration) was applied to the calibrated visibilities to generate final VLBI images. 
We performed a ridge-line analysis and Gaussian model-fits on the final jet image to derive the jet parameters. 
In our data, the presence of NOEMA improves the image sensitivity by a factor of 2.5. The jet shows a clear wiggling structure within 0.4\,mas from the core. 
Our ridge-line analysis suggests the presence of a helical jet structure (i.e., a sinusoidal pattern).
Six circular Gaussian components were fitted to the inner jet region. 
We estimated an apparent brightness temperature of $\sim$3\,$\times$\,10$^{12}$\,K in the two innermost components. 
They are likely to be highly boosted by relativistic beaming effect.
We find four significant polarized knots in the jet. Interestingly, two of them are located in the core region. Finally, we suggest a number of physical scenarios to interpret our results.
}


\maketitle
%

\section{Introduction}
\label{intro}
At millimeter wavelengths, very long baseline interferometry (VLBI) provides a sub-milliarcsecond scale (sub-mas) resolution, which facilitates the study of the jet-launching in the central region of active galactic nuclei (AGN) \citep{krichbaum1998}.
Powerful radio jets launched in the vicinity of supermassive black holes (SMBHs) have been the primary target of VLBI studies owing to their compact, sub-pc scale structures, and bright non-thermal emission across the entire electromagnetic spectrum \citep{zensus1997, blandford2019}. 
However, our understanding of the jet launching region remains limited due to insufficient angular resolution and jet opacity \citep[i.e., the so-called core shift;][]{lobanov1998}.
The inner jet region is thought to be optically thick at radio due to synchrotron self-absorption. By observing at higher frequencies (e.g., 86 GHz), we can peer deeper into the near-horizon scales of the jet structures.

The Global mm-VLBI Array\footnote{https://www3.mpifr-bonn.mpg.de/div/vlbi/globalmm/} (GMVA) allows us to explore the inner subpc-scale jets in more detail \citep[e.g.,][]{baczko2016, boccardi2016, hodgson2017, kim2019, casadio2021, paraschos2022}. The GMVA operates at 86\,GHz (3.5\,mm) and is capable of reaching a very high angular resolution of $\sim$50\,micro-arcsecond ($\mu$as).
The presence of sensitive and large antennas in the array is the key for better image fidelity and thus high-quality scientific output. 
The NOrthern Extended Millimetre Array\footnote{https://iram-institute.org/observatories/noema/} (NOEMA) is an interferometer that had 11 antennas at the time of the observations reported here (12 today) that can either be operated in local interferometer mode or phased array mode. The 2SB dual polarization receiver bands of $4\,\times\,7744$\,MHz processed by the PolyFiX correlator allow phased array operation over the full bandwidth. Based on the measurements in April 2021, NOEMA has a system equivalent flux density (SEFD) of 150--250\,Jy with a typical system temperature (Tsys) of $\sim$80\,Kelvin (K) at 86 GHz. This is a major upgrade from the old six-element Plateau de Bure Interferometer (PdBI) that has participated in GMVA observations since 2003. NOEMA has twice the collecting area and four times the spectral bandwidth of the PdBI. In the phased array mode, data rates have passed from maximum 1 GBit/s (PdBI, correlator limited) to 64 GBits/s (NOEMA, recorder limited; the correlator can do up to 128 GBits/s). In April 2021, NOEMA joined regular GMVA observations for the first time.

\begin{figure}[!t]
\centering
\includegraphics[angle=0, width=88mm, height=50mm]{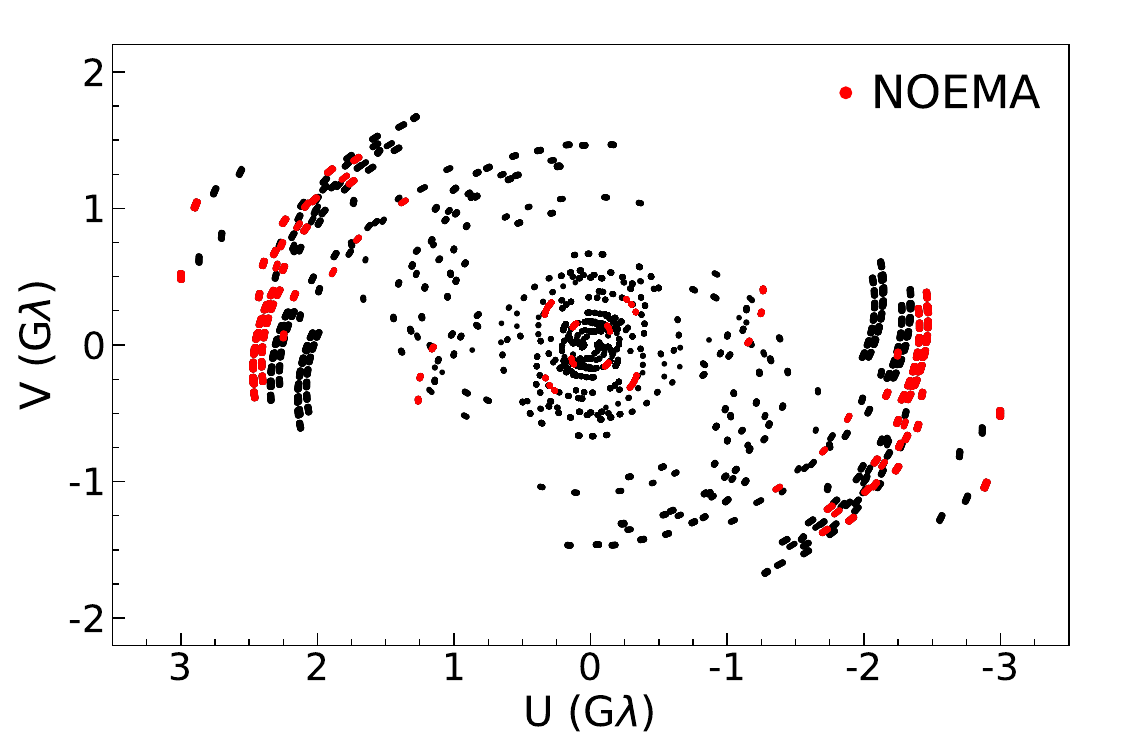} \
\includegraphics[angle=0, width=88mm, height=50mm]{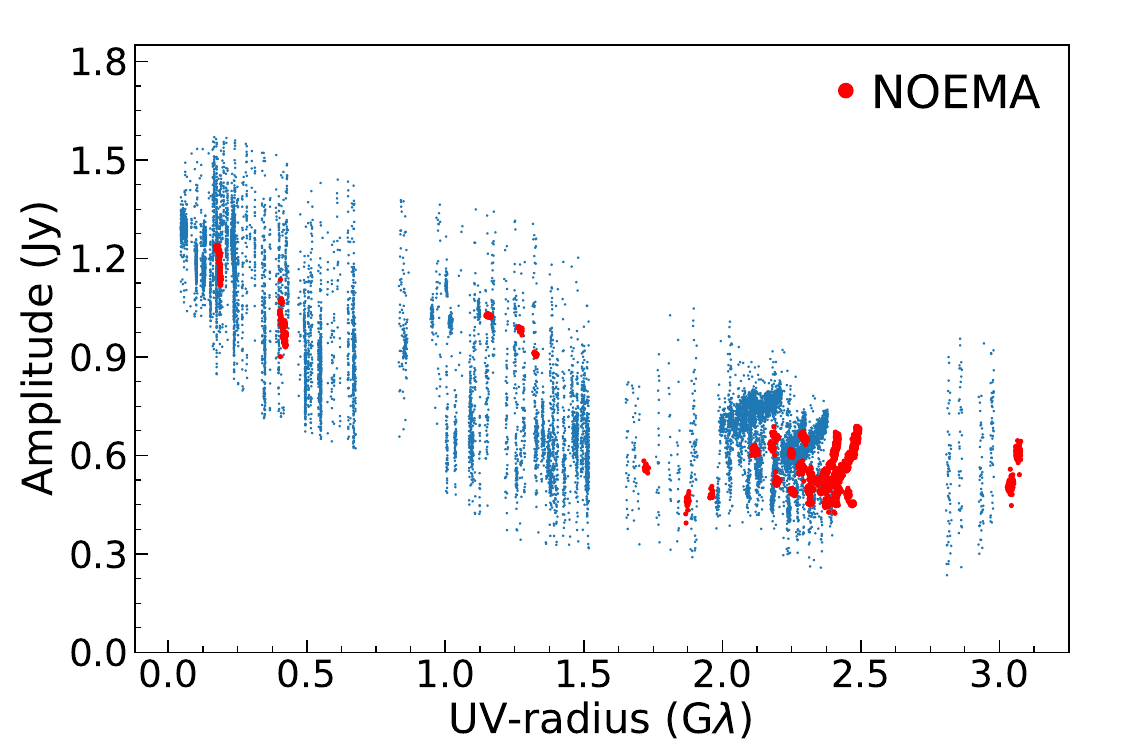} \
\includegraphics[angle=0, width=88mm, height=50mm]{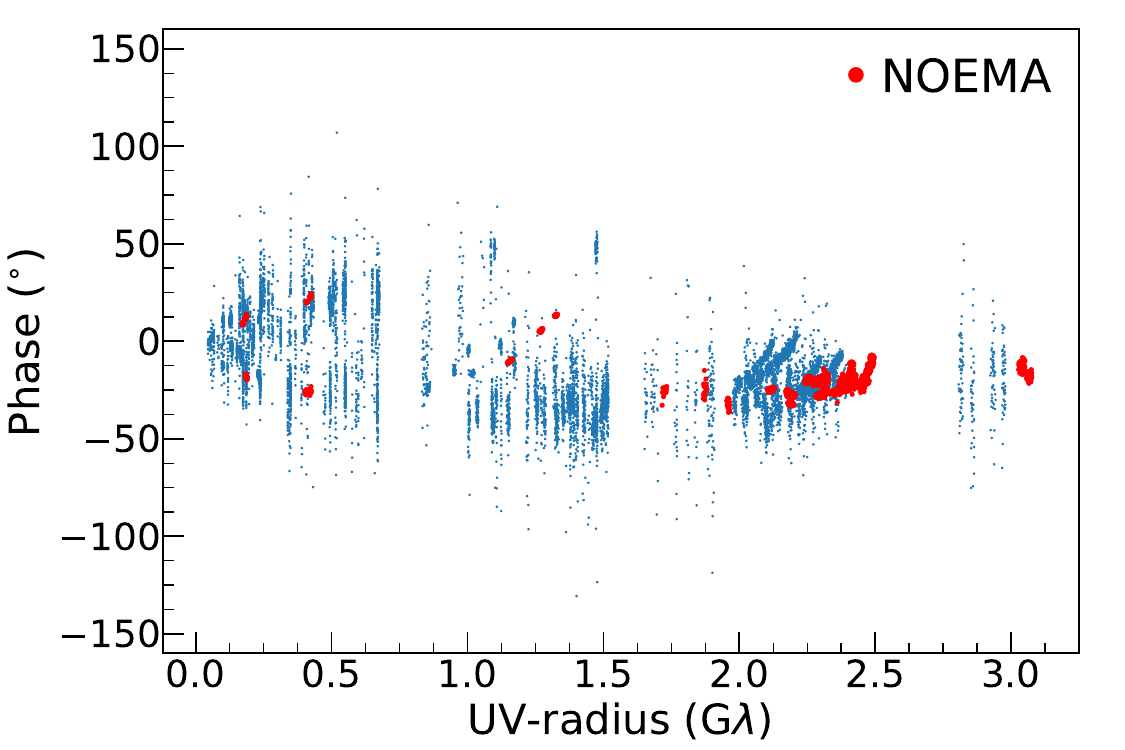}
\caption{\textsl{uv}-coverage (\textsl{top}), visibility amplitude (\textsl{middle}), and phase (\textsl{bottom}) of the fully calibrated data. The NOEMA baselines are indicated by red color. 
For uniform weighting, a beam size of 0.165\,$\times$\,0.037\,mas at $-$4.72$^{\circ}$ is obtained.
}
\label{fig:uvcov}
\end{figure}

BL\,Lacertae (BL\,Lac; 2200$+$420) is a prototypical source of the BL\,Lac class of objects and it is a nearby AGN: its redshift ($z$) is $z$\,$\sim$\,0.0686 \citep{vermeulen1995}, corresponding to an image scale of 1.3\,pc/mas (assuming $H_{0}$ = 71 km~Mpc~s$^{-1}$, $\Omega_{\Lambda}$ = 0.73, and $\Omega_{m}$ = 0.27).
The jet of BL\,Lac is a strong $\gamma$-ray emitter that is variable with time and has been linked to (sub)pc-scale jet activity such as the propagation of moving blobs and shocks \citep{marscher2008, wehrle2016}.
The morphology of the jet is known to be very complicated, with prominent bending or wiggling structures. 
The origin of such behavior is still a matter of debate. 
Possible interpretations invoke for instance, a helical jet model \citep{denn2000}, jet precession \citep{stirling2003}, MHD waves \citep{cohen2015}, and kink instabilities \citep{jorstad2022}. 
It was recently reported \citep[][]{push2023} that the electric vector position angles (EVPAs) of the mas-scale jet are well-aligned with the local jet axis.
At higher frequencies $\geq$\,43\,GHz, however, the appearance of bright polarized emission seems erratic in the upstream region around the radio core \citep[e.g.,][]{marscher2008, wehrle2016}. 
These signatures favor the presence of 
the dominant toroidal magnetic fields in the jet and multiple emitting regions in the unresolved upstream jet region \citep{hovatta2012}.


\section{GMVA observations in April 2021}
\label{sec:data}
In this work, we focus on the analysis of BL\,Lac that was observed as a calibrator 
in the project \texttt{MB018A} (PI: B. Boccardi, main target: NGC\,315). 
The observations were performed on 24 April 2021. 
The duration of the observation was $\sim$15\,hrs and the actual on-source time on BL\,Lac was $\sim$1.6\,hrs; in the case of NOEMA, it was $\sim$1.1\,hrs. The number of phased antennas for NOEMA was 10 in this experiment.
The data consists of 8 IFs over 86.1--86.4\,GHz with a bandwidth of 64\,MHz for each. Overall, 17 radio telescopes participated in the observation: Effelsberg (EF), Onsala (ON), Mets\"{a}hovi (MH), NOEMA (NN), Pico Veleta (PV; the IRAM 30m telescope), and Greenland (GLT), along with eight VLBA antennas and three KVN antennas. 
We used the novel \texttt{rPICARD} pipeline \citep{janssen2019} for a reproducible data reduction and \texttt{GPCAL} \citep{park2021a} for polarization calibration. 
In Appendix~\ref{sec:calim}, we summarize the detailed steps of the data analysis. We also compare the results between the standard data analysis done in \texttt{AIPS} \citep{greisen2003} with the new analysis pipeline \texttt{rPICARD}, which relies on \texttt{CASA} \citep{casa2022}.

\begin{figure}[t]
\centering
\includegraphics[angle=0, width=\columnwidth, height=5.5cm]{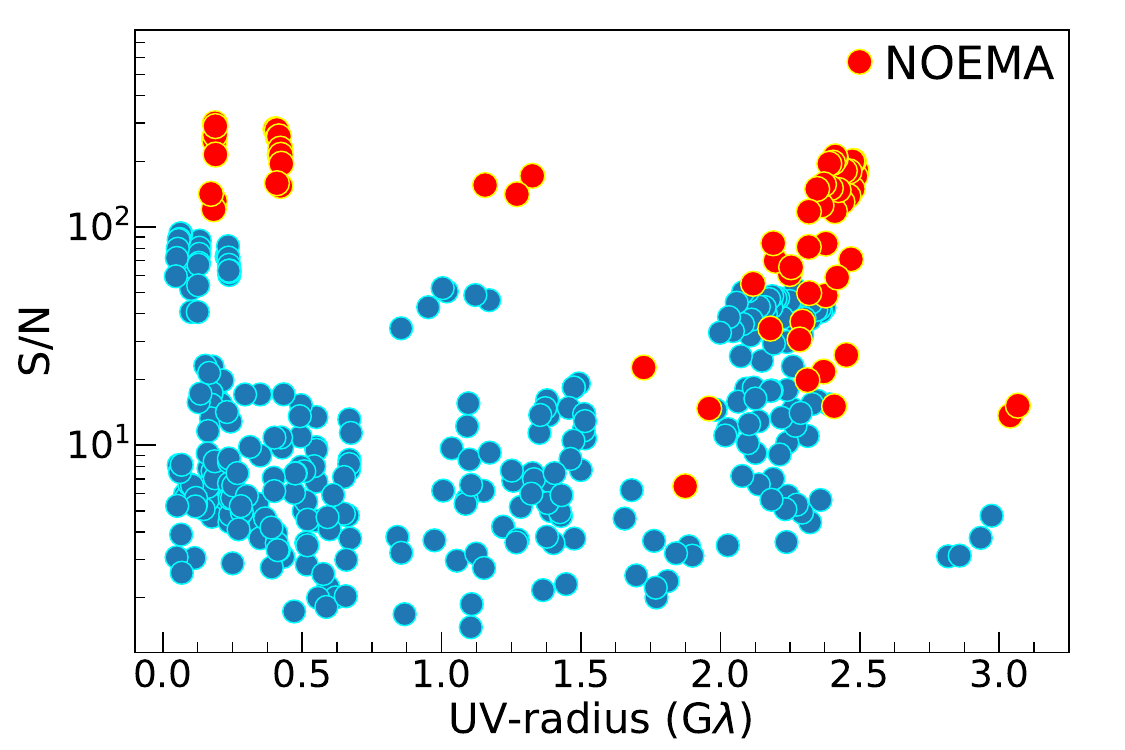}
\caption{S/N plot of the calibrated data. The data were scan- and frequency-averaged in this plot.
}
\label{fig:thesnr}
\end{figure}

\begin{table}[t]
\caption{Residual rms noise levels ($\sigma_{\rm rms}$) of the final image.}
\label{tab:rms}
\centering 
\begin{tabular}{l @{\hspace{15mm}} c c r}
\hline      
ID & $\sigma_{\rm rms}$\,$^{a}$ & Ratio\,$^{b}$ & SEFD\,$^{c}$  \\
  &  (mJy/beam)  &    &  (Jy)  \\
\hline
All antennas  &  2.35  &  -  &  -  \\
w/o NOEMA\,$^{d}$  &  5.92  &  2.5  &  $\sim$190  \\
\hline
\multicolumn{4}{l}{$^a$ Measured at the map center (a region of 4\,mas\,$\times$\,4\,mas).}\\
\multicolumn{4}{l}{$^b$ Ratio of the noise level to the one with all antennas.}\\
\multicolumn{4}{l}{$^c$ Overall SEFDs of the station in the data.}\\
\multicolumn{4}{l}{$^d$ All antennas but without NOEMA.}\\
\end{tabular}
\end{table}

Figure~\ref{fig:uvcov} shows the fully calibrated visibilities of the data. 
The NOEMA baselines are marked in red color. The shortest baselines include NOEMA--Effelsberg (660\,km), the longest baseline is NOEMA--VLBA\_MK (Hawaii, 10670\,km).
From simple visual inspection of the visibility plots, we can easily find that the groups of the NOEMA visibilities are very compact without large scattering in amplitude/phase, compared to other baselines in the observation. 
This indicates that they have smaller visibility errors and higher signal-to-noise ratio (S/N).
Indeed, we find that the overall S/Ns of the NOEMA datasets are significantly higher than the others by a factor of at least $\geq$\,3 (see Figure~\ref{fig:thesnr}).
In addition to that, we refer to Table~\ref{tab:rms} for the impact of NOEMA on imaging sensitivity.

\section{Results}
\label{sec:result}

\subsection{Total intensity structure of the jet}
\label{sec:ridge}
\begin{figure}[t]
\centering
\includegraphics[angle=0, width=\columnwidth, keepaspectratio]{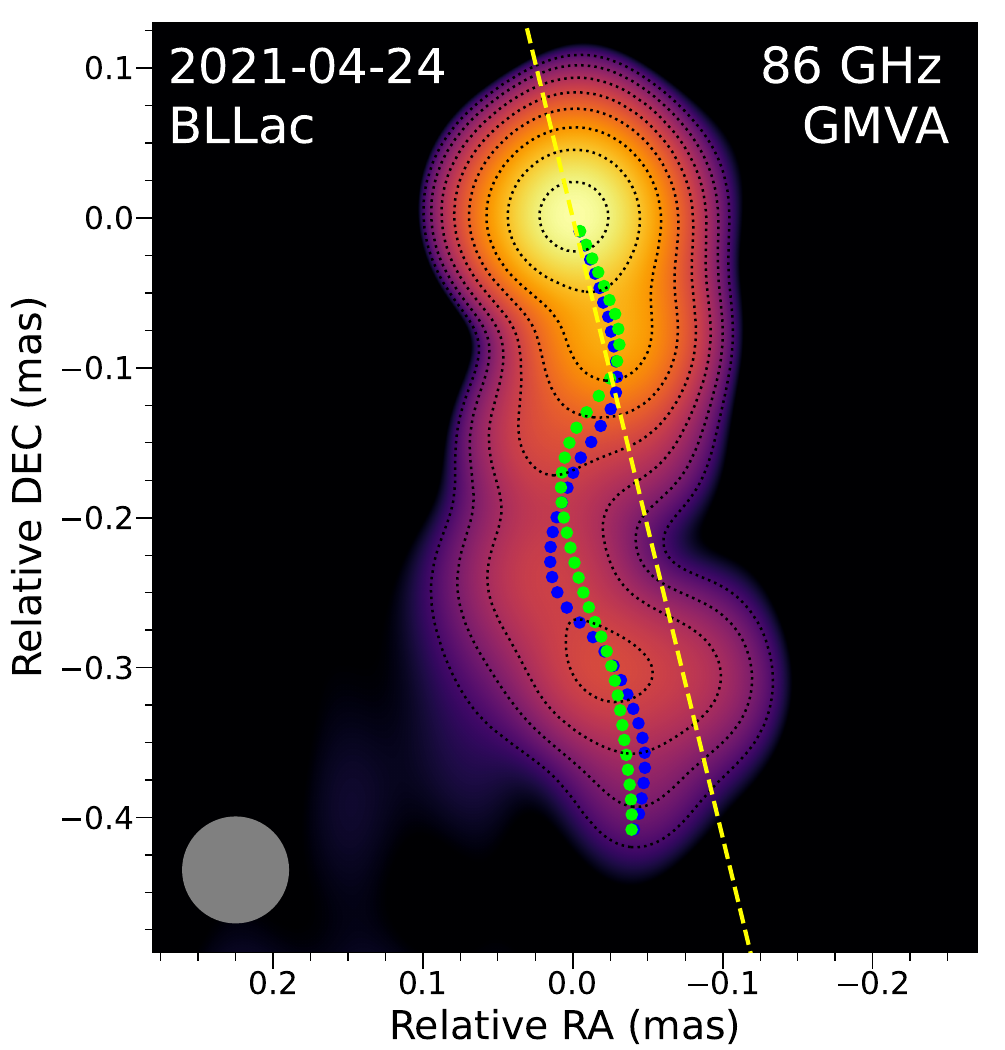}
\caption{Jet of BL\,Lac at 86\,GHz observed by the GMVA on 24 April 2021. The color scale and contours show the total intensity structure of the jet (natural weighting). The contours increase by a factor of 2 from 0.6\,\% to 76.8\,\% of the map peak ($\sim$0.77\,Jy/beam). The gray circle at bottom-left denotes the restoring circular Gaussian beam of 0.07\,mas (= 70\,$\mu$as). The green and blue dots indicate the ridge lines measured with the model-fit and CLEAN maps, respectively. The yellow dashed line points to the direction of the known jet component $B15$ presented in \citet{weaver2022}.
}
\label{fig:clnmap}
\end{figure}

Figure~\ref{fig:clnmap} shows a final CLEAN image of the jet obtained from the 2021 GMVA observations. The overall data quality of the VLBA was not the best in the observation and we were only able to image the jet down to $\sim$0.4\,mas from the core due to limited sensitivity. 
In our maps, the source structure is spatially resolved. 
We find the common core-jet morphology \citep[e.g.,][]{jorstad2017}, but, interestingly, it displays a clear wiggling (i.e., a multiple-bending feature) structure 
determined by ridge line analysis (see below). 
Such behavior has already previously been reported in the source  \citep[e.g.,][]{cohen2015}, but at parsec scales on the sky plane. There is a persistent, bright spot in the jet at 
a core separation of $r$\,=\,1.5\,mas; \citet{weaver2022} identified this component as $B15$. The position angle of $B15$ (i.e., $-$166.4$^{\circ}$, CW from 0$^{\circ}$ at north to $-$180$^{\circ}$ at south) and the jet flow shown in Figure~\ref{fig:clnmap} (i.e., $-$172$^{\circ}$ on average) are different which suggests a misalignment of the jet axis between subpc and pc scales.

\begin{figure}[t]
\centering
\includegraphics[angle=0, width=\columnwidth, keepaspectratio]{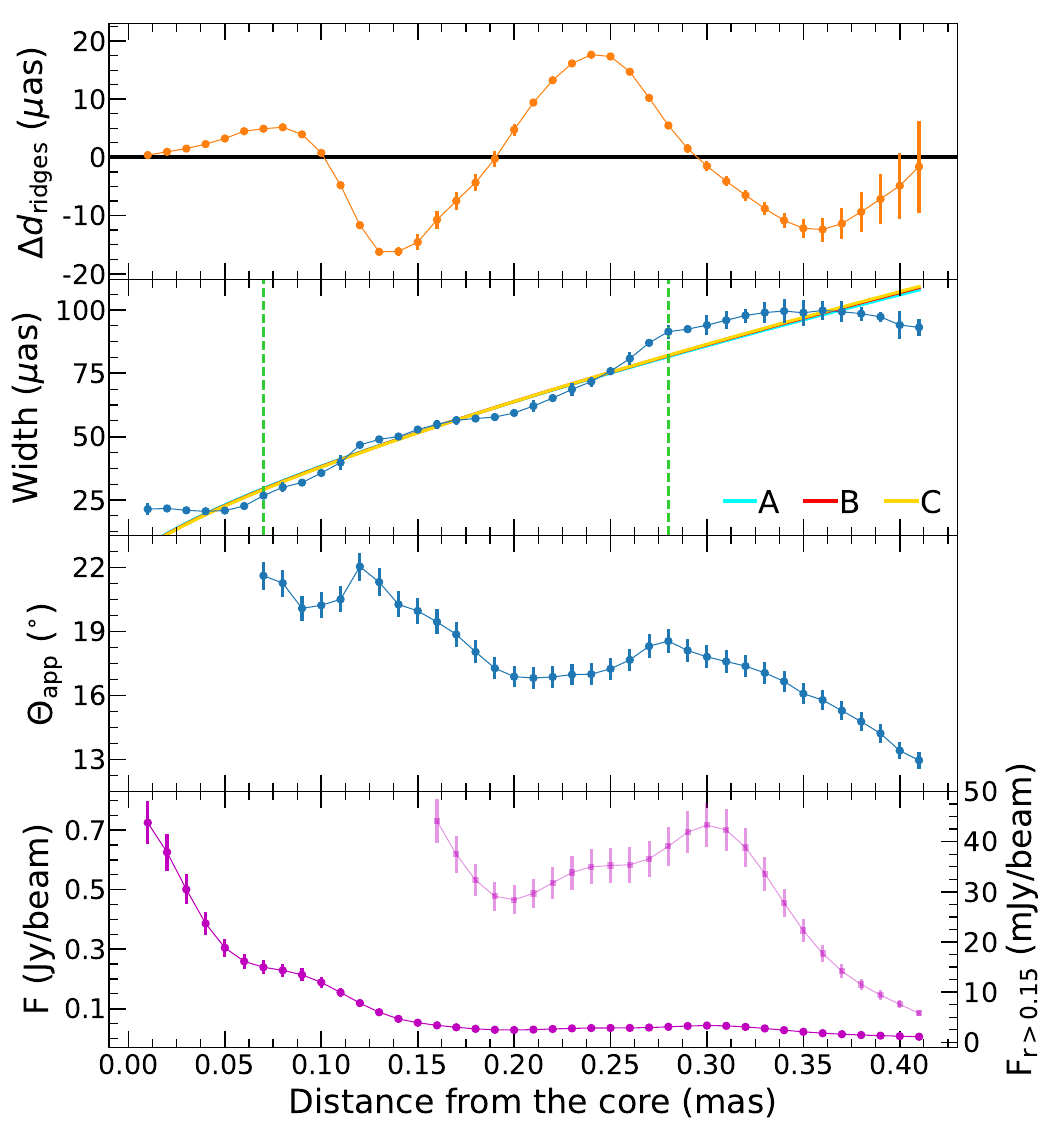}
\caption{Distance between the two ridge lines at every 0.01\,mas from the core ($\Delta d_{\rm ridges}$), transverse jet width ($W$), apparent jet opening angle ($\Theta_{\rm app}$), and a profile of flux density as a function of the distance from the core ($r$), shown from top to bottom. 
A simple power-law form is fitted to the width profile in three different areas divided by $r$ = 0.07\,mas and 0.28\,mas to take into account any flattening feature resulting from resolution limits and low S/N.
The opening angles are considered only when $r$\,$\geq$\,0.07\,mas. The outer part of the emission profile (i.e., $r$\,$>$\,0.15\,mas) is enlarged separately. More details of the measurements can be found in Appendix~\ref{sec:rganal}.
}
\label{fig:jetpara}
\end{figure}

We also fit the data with circular Gaussian components (see Appendix~\ref{sec:ccs}, for their physical parameters). In both the CLEAN and model-fit images, the overall jet morphology are consistent with each other. 
From visual inspection, however, they look slightly different. 
To further examine the differences and measure the jet properties, we draw jet ridge lines by using the two images (see Appendix~\ref{sec:rganal}, for details of the analysis). Figures~\ref{fig:clnmap} and \ref{fig:jetpara} show the results (hereafter, CN ridge for the CLEAN jet and MD ridge for the model-fit jet). As evident from the comparison, the wiggling pattern appears more prominent in the CN ridge than the MD ridge. We consider that the model-fit image describes the jet flow with a number of Gaussian components of certain sizes focused on brightest regions along the jet downstream, whereas the CLEAN process models the jet image with a set of many delta-components (i.e., point sources) that reflects faint emission regions more effectively. Hence, any resolved weak source structures can be better highlighted in the CLEAN image, which result in the discrepancy. This suggests that the CN ridge shows the jet structure in more detail. The reduced Chi-squared values of the CLEAN and model-fit images (i.e., $\sim$1.2 and $\sim$1.8, respectively) also support this idea. In this context, the MD ridge could be considered as a line that just connects the representative (or brightest) regions of the flow.

The jet of BL\,Lac is known for its complex, wiggling structure that varies over time \citep[e.g.,][]{cohen2015}. Such behavior can be clearly seen in the image with the CN ridge line. The CN ridge shows multiple bending points at $\sim$0.11\,mas, $\sim$0.23\,mas, and $\sim$0.37\,mas from the core. In addition to that, interestingly, the CN ridge line looks rotating around the MD ridge line. The observed features lead us to consider a helical jet structure that theoretical simulations have  predicted \citep[e.g.,][]{meier2001, mizuno2011}.
In Figure~\ref{fig:jetpara}, we present several jet parameters estimated by the ridge lines. 
Overall, the distance between the two ridge lines ($\Delta d_{\rm ridges}$) is below the level of angular resolution of the data and the sizes of the Gaussian components. This prevents us from making an argument quantitatively. However, there is an obvious systematic behavior in $\Delta d_{\rm ridges}$ that is sinusoidal along the downstream. This might be a hint of the helical jet structure, rather than just a coincidence. 
We fit a simple power-law to the measured jet width ($W$): $W(r) = a r^{k}$ where $a$ and $k$ being free parameters. 
To consider some of the known issues with resolution limit near the core and low S/N at the tail part of the jet, we checked the fitting results in three different areas: (A) the whole jet ridge line, (B) $r$\,$\geq$\,0.07\,mas, and (C) between 0.07\,mas and 0.28\,mas. 
The best-fit parameters of the fits are on average: $a$ = 0.21$\pm$0.02 and $k$ = 0.74$\pm$0.05 (with 1$\sigma$ uncertainty); the three fits resulted in the index $k$ of 0.73--0.75, thus remaining consistent with respect to each other within the errorbars.
Previous works measured the collimation profile of the jet at various observing frequencies with different approaches: stacked VLBI images \citep{kovalev2020, casadio2021} and model-fit components of single-epoch images over a long-term period \citep{burd2022}. Their results on the inner jet region span a wide range; $k$ spans the interval 0.5--1.2, larger at higher frequencies. Our estimate (i.e., $k$\,$\sim$\,0.74) is in the middle of this range. This tendency suggests that the jet structure at subpc scales is highly complicated. At higher observing frequencies, the source is better resolved and thus the structural complexity (or time-variable local jet axis to left and right, which is probably magnified by projection effects) could better be reflected horizontally on the image plane. 
Then, the stacking of multi-epoch, high-resolution images of such an extreme jet could make its width much wider than a single-epoch image. 
\citet{lister2013} considered that in a conical jet model, a single-epoch image may show us a part of the jet rather than the entire jet cross section. However, it is unclear whether the subpc-scale jet of BL\,Lac is conical ($k$\,$\sim$\,1.0) or parabolic ($k$\,$\sim$\,0.5), considering those different estimates found by the previous works.
In our future works, we plan to address this issue in more detail by using VLBI observations of BL\,Lac with the Event Horizon Telescope (EHT) at 230\,GHz.
We also note that the overall pattern of our CN ridge line already looks quite similar to the ridge line presented in \citet{casadio2021} that was obtained from a stacked GMVA image of BL\,Lac at 86\,GHz (i.e., their Figure~2).

Overall, the apparent jet opening angle ($\Theta_{\rm app}$) decreases with the distance from the core. Using a single-epoch observation, \citet{hada2018} and \citet{park2021b} also found the same trend in the parabolic jet regions of the Narrow-line Seyfert 1 Galaxy 1H0323+342 and the radio galaxy NGC\,315, respectively \citep[but see also][for opposite case in a blazar]{ros2020}. 
The apparent opening angle can be varied due to changes in the intrinsic opening angle ($\Theta_{\rm int}$) and/or the jet viewing angle ($\theta_{\rm jet}$) which is attributed to the projection effect \citep[e.g.,][]{homan2002}: $\Theta_{\rm app}$ = $\Theta_{\rm int}$\,/\,sin($\theta_{\rm jet}$). We find two local peaks in $\Theta_{\rm app}$ at $r \sim$ 0.12\,mas and 0.28\,mas. The first one is accompanied by a short dip in $W$ at $\sim$0.1\,mas and a strong emitting component around it. Thus, it might probably be caused by changes in the pressure of the plasma flow. 
In the case of the second one, we suspect changes in $\theta_{\rm jet}$ values around that region. A continuous, increasing pattern in flux density where 0.2\,$<$\,$r$\,$<$\,0.3, coincides with the second peak. This might be caused by smaller $\theta_{\rm jet}$ values in this region.

\subsection{Linearly polarized emission}

\begin{figure}[t]
\centering
\includegraphics[angle=0, width=\columnwidth, keepaspectratio]{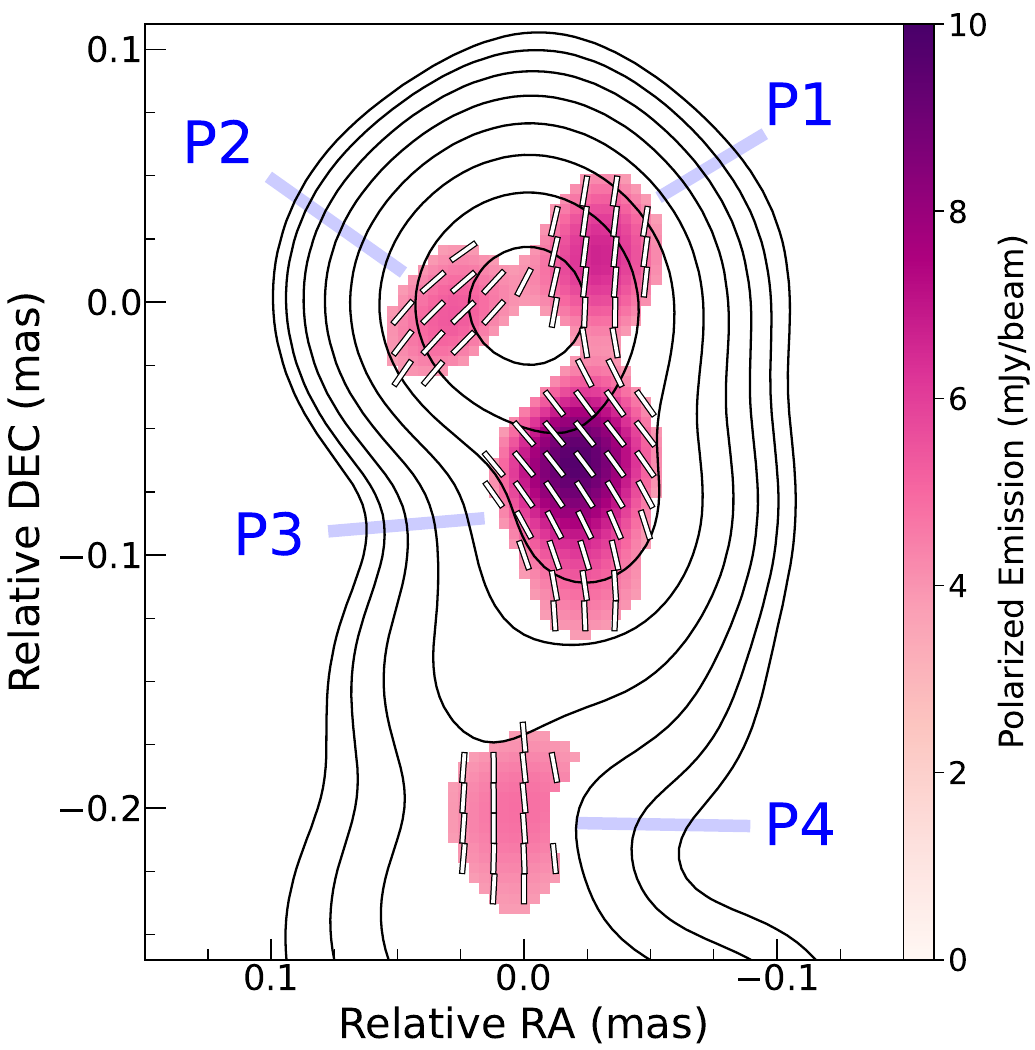}
\caption{
Same as Figure~\ref{fig:clnmap}, but the distribution of linearly polarized emission (color scale) is overlaid onto the total intensity structure (contours). The white line segments represent EVPAs of polarized knots. Pixel cutoffs of 10$\sigma_{\rm I}$ for total intensity and 4$\sigma_{\rm P}$ for the polarization were used in this map with natural weighting. The polarized knots are labeled by P1--4. The EVPAs are not Faraday-corrected.
}
\label{fig:mainpol}
\end{figure}

Figure~\ref{fig:mainpol} shows the distribution of linearly polarized emission of the jet. 
We found four significant polarized knots within $r$\,=\,0.25\,mas from the core. 
The fractional polarization of the four knots are roughly: $\sim$2\,\% (P1), $\sim$1\,\% (P2), $\sim$4\,\% (P3), and $\sim$17\,\% (P4); measured at their central regions. 
We refer to Appendix~\ref{sec:polsigtest} for details of the polarization detection and parameters of the four emission regions.
In the case of the core (or the map peak), it is around 0.5\,\%. The overall EVPAs in the jet are parallel to the local jet direction; i.e., the P3 and P4 knots aptly describe the flow axis. However, the two knots near the core in the upstream region (i.e., P1 \& P2) seem to be different. The angular resolution of our data is insufficient to resolve this region and, thus, the detailed jet axis and structure are hidden.
\citet{push2023} presented a stacked polarization image of the jet of BL\,Lac at 15\,GHz, by using 139 VLBA images of the source spanning over 20\,years. In their image, the jet EVPAs are well aligned with the flow down to, for instance, $\sim$6\,mas from the core. The downstream knots in our image (i.e., P3 \& P4) confirm that such behavior persists up to about 0.05--0.1\,mas from the core. This is also consistent with \citet{rani2016}. 
Faraday rotation is known to be very weak in the jet on parsec scales \citep{hovatta2012}; however, it is unclear at the upstream region where the radio core appears. 
The EVPAs of P3 and P4 suggest the presence of strong toroidal magnetic fields in this subpc-scale region of the jet and probably also further downstream (i.e., on pc scales). 
The stacked jet image of \citet{push2023} supports this; they used 139 VLBA maps of the jet at 15\,GHz, which were not Faraday-corrected. 
This might be an indication of the presence of a helical magnetic field configuration in this inner jet \citep[e.g.,][]{nakamura2010}.

At lower frequencies, in general, the cores of blazar jets show a single polarized knot that has no significant displacement from it, except when a strong disturbance propagates down the flow 
(e.g., \citealt{kim2018}; see also \citealt{issaoun2022, jorstad2023}, for VLBI images of blazar jets with the extended core polarization structures at 230\,GHz). 
In our image, there are two polarized knots P1 and P2, near the core with a slight offset from the map center (i.e., $\sim$35\,$\mu$as for P1 and $\sim$30\,$\mu$as for P2). This is an unusual case. However, it seems that these features are also present in VLBI images of some previous works \citep[i.e.,][]{marscher2008, rani2016, casadio2017}; in particular, both knots can be clearly seen in \citet{casadio2017}. In addition to that, we checked a near-in-time (i.e., 5 April 2021) 86\,GHz VLBA image of the jet offered by the Boston University group\footnote{https://www.bu.edu/blazars/index.html} (BU) (see Appendix~\ref{app:bu3mm}). Although this data is resolution-limited, we see hints of the P1 and P2 knots around the core with the similar EVPA patterns in the VLBA image.
It is unclear what those features indicate without more detailed information on the core region (e.g., intrinsic EVPAs and opacity). \citet{hovatta2012} could not measure rotation measure (RM) in the core of BL\,Lac due to the blending effect of multiple components; they employed the $\lambda^{2}$-fit to find RM values at cm-wavelengths. Indeed, our image shows such feature (i.e., P1 \& P2). At higher frequencies (i.e., 15--43\,GHz), \citet{gomez2016} found RM values of a few thousands of rad/m$^{2}$ (up to 3000\,rad/m$^{2}$) around the core. Thus, the presence of strong Faraday rotation cannot be ruled out in this upstream region of the jet and high resolution multi-frequency mm-VLBI observations at, for instance, $>$\,43\,GHz are required to shed light on this region. It is worthwhile noting that just three\,days after our observation, there was the strongest $\gamma$-ray outburst that is unprecedented in this source over the past $\sim$15\,years (see Appendix~\ref{sec:fermi}).

\section{Discussion}

\subsection{On the geometry of the subpc-scale jet}
\label{dis:geo}

In this section, we suggest a model of the jet geometry in a qualitative way, considering the observed structural features. Our findings favor the hypothesis of helical jet structure \citep[e.g.,][]{meyer2018}. Although it is resolution-limited, the systematic sinusoidal pattern in $\Delta d_{\rm ridges}$ looks obvious and this can naturally be explained by the helical jet structure.
In this view, the two ridge lines might actually show us the helical jet streamline (i.e., the CN ridge) and the changing axis of the helix (i.e., the MD ridge), respectively. We found three bending points in our analysis. The jet is assumed to have a small viewing angle (e.g., $\theta_{\rm jet}$\,$<$\,10$^{\circ}$) and flow downstream helically. 
Thus, there could be changes in the direction of the actual flow (i.e., $\theta_{\rm streamline}$) around those bending regions, with respect to the line of sight 
(i.e., decreasing and increasing). 
We notice that the flux density increased between 0.2\,mas and 0.3\,mas in $r$. An enhanced beaming effect could be responsible for the rise \citep[e.g.,][]{raiteri2017}, thus meaning that the flow (i.e., streamline) was approaching (or better aligned) to the observer in that region. If this is the case, we expect that the jet was initially approaching, and then it turned into receding (by the helical motion) at around $r$\,$\sim$\,0.1\,mas. Beyond $\sim$0.4\,mas, the jet disappears quickly and this can be found frequently in previous works \citep[e.g.,][]{casadio2021}. This might be due to a relatively large change in the axis of the jet helix $\theta_{\rm jet}$ towards a large angle.

Details of the upstream region (i.e., $r$\,$<$\,0.1\,mas) are unclear due to the limited angular resolution. 
We found the observed brightness temperature (T$_{\rm b,obs}$) of $\sim$3\,$\times$\,10$^{12}$\,K in this upstream region ($C0$ \& $C1$). T$_{\rm b,obs}$ can be inferred from the Doppler factor ($\delta$) and the intrinsic brightness temperature (T$_{\rm b,int}$) as T$_{\rm b,obs}$ = T$_{\rm b,int}\,\delta$\,/\,(1 + $z$), where $z$ and $\delta$ being redshift and $\delta$ = [$\Gamma$($1\,-\,\beta\,{\rm cos}\,\theta_{\rm jet}$)]$^{-1}$, respectively; $\Gamma$ is the Lorentz factor ($\Gamma$ = 1\,/$\sqrt{1 - \beta^{2}}$) and $\beta$ is the jet speed in units of the speed of light. 
Using the typical $\delta$ values of the jet reported by \citet{jorstad2017} and \citet{weaver2022}: namely, 7.5 and 4.3, respectively, we estimate T$_{\rm b,int}$ of $\sim$4\,$\times$\,10$^{11}$\,K for $\delta$ = 7.5 and $\sim$7\,$\times$\,10$^{11}$\,K for $\delta$ = 4.3. 
These estimates are higher than the equipartition limit (T$_{\rm b,eq}$) of $\sim$5\,$\times$\,10$^{10}$\,K \citep{rhead1994} that assumes equipartition in energy density between the magnetic field and emitting particles; \citet{liodakis2018} recently reported T$_{\rm b,eq}$ to be $\sim$2.8\,$\times$\,10$^{11}$\,K (i.e., the maximum T$_{\rm b,int}$ measured independently of the assumption of equipartition). This assumption has been common practice in the relevant studies \citep{homan2021}.
In this context, we suggest two scenarios to interpret our T$_{\rm b,obs}$ values found in $C0$ and $C1$. Firstly, the emission regions might be particle-dominated during the time of our observation (i.e., the flaring state). Secondly, if we follow the assumption of equipartition (i.e., T$_{\rm b,int}$\,$\sim$\,T$_{\rm b,eq}$), then $\delta$ should be $\sim$64 for \citet{rhead1994} and $\sim$11.5 for \citet{liodakis2018}. This indicates the presence of highly enhanced Doppler boosting in these regions.
For $C1$, we consider that opacity is lower than $C0$ \citep[see e.g.,][for a spectral index map of the jet]{gomez2016}, thus suggesting lower $\delta$ values than the above estimates (e.g., $\sim$32 and $\sim$8, respectively). Using a GMVA survey at 86\,GHz, \citet{nair2019} presented a statistics on T$_{\rm b,obs}$ for a large source sample. Interestingly, the T$_{\rm b,obs}$ values we found in the upstream region, are about two times larger than the highest value of \citet{nair2019}, namely: T$_{\rm b,obs}$\,$\sim$\,1.3\,$\times$\,10$^{12}$\,K. Thus, the upstream region of the jet is likely more aligned to the line of sight with smaller viewing angle. We also note that if the jet is viewed at a very small viewing angle (i.e., close to $\sim$0$^{\circ}$), then the projected EVPAs can be less dependent on the jet orientation \citep[e.g.,][]{jorstad2003}, which might be reflected through the P1 and P2 knots in the core region. Lastly, we further add a possibility of the recollimation-shock (RCS) scenario for the core polarization, considering the structured linear polarization of RCS \citep[see e.g.,][]{marscher2016}.

\subsection{Coincidence with the strongest \texorpdfstring{$\gamma$}{g}-ray outburst}
\label{dis:gam}

Luckily, BL\,Lac was in a peaking stage of its historically huge $\gamma$-ray flare and it peaked just three days after our observation. The enhanced jet activity at $r$\,$<$\,0.1\,mas in terms of polarized emission and T$_{\rm b,obs}$, is a strong evidence that the origin of the $\gamma$-ray event was in this upstream region. Interestingly, the source was also flaring at 230\,GHz in April/May 2021, based on the Submillimeter Array (SMA) database\footnote{http://sma1.sma.hawaii.edu/callist/callist.html}; due to limited cadence, it is unclear when the radio flare peaked exactly, but it seems to be late-April or early-May 2021 with a level of $\sim$6\,Jy.
Using a\ spectral energy distribution (SED) modeling, \citet{sahakyan2023} found that size of the emitting region tends to be smaller during a flaring state, which in our case, favors $C1$ as the main production site of the $\gamma$-ray event. 
However, the core region (i.e., $C0$) could also be responsible for the $\gamma$-ray emission. \citet{macdonald2015} proposed the ring of fire model that assumes the presence of a synchrotron-emitting ring of electrons providing soft seed photons for the inverse-Compton process in the jet. The two polarized knots (i.e., P1 \& P2) might indicate the ring that is a shocked portion of the jet sheath. Indeed, the transverse emission profiles in \citet{macdonald2015} (i.e., their Figures~7 \& 8) show a similar pattern to the core region of our image with P1 and P2. We assume that the strongest $\gamma$-ray outburst of BL\,Lac was not an orphan event owing to the presence of a near-in-time radio counterpart at 230\,GHz (SMA). 
This could be an indication that P1 and P2 are located downstream of the core \citep[see][for an orphan $\gamma$-ray flare with the ring located upstream of the core]{macdonald2015}. Then, this scenario suggests that the $\gamma$-ray outburst is originated near the $C0$ region, probably when a moving feature \citep[e.g.,][]{kim2022} with highly accelerated electrons passes through the ring area.
Alternatively, one may also consider relativistic magnetic reconnection as the underlying physical process of the $\gamma$-ray outburst. The reconnection dissipates magnetic energy and accelerate particles efficiently in the jets. \citet{zhang2022} reported that a fast, strong $\gamma$-ray flare can be produced by plasmoids (or mini-jets) in the reconnection region. Such plasmoids are thought to show enhanced polarized emission and a large swing in polarization angle caused by plasmoid mergers, which might be reflected in our image as the polarized features in the core region \citep[see also][for discussion on a link between magnetic reconnection and kink instability in the jet of BL\,Lac]{jorstad2022}.

\section{Summary}
\label{sec:fin}

We have analyzed the first GMVA data, where the upgraded NOEMA interferometer joins as a phased array. From the observations taken in April 2021, we studied the BL\,Lac data in this work.
Fortunately, the observation coincides with the strongest $\gamma$-ray flare of this source, which makes an examination of the data all the more interesting. The participation of NOEMA improved the imaging sensitivity largely (i.e., by a factor of $\sim$2.5 in our data) and the capability of detecting small-scale source features. The advanced calibration pipelines worked on the data nicely. In particular, we found that the visibility phase errors can be reduced significantly with \texttt{rPICARD}, compared to the conventional manual calibration with \texttt{AIPS}.
Our GMVA image of the jet of BL\,Lac gives us hints of the complex nature of the jet at subpc scales from the central engine, such as the wiggling structure, unusual polarization signatures, and enhanced brightness temperatures. We suggest a number of scenarios to explain those interesting features in the presence of helical jet structure.
However, we expect that the jet of BL\,Lac is highly variable with time. Thus, a monitoring of the source with high-resolution ($<$\,50\,$\mu$as) mm-VLBI array would be highly useful for shedding more light on the nature of the subpc-scale jet of BL\,Lac.

\begin{acknowledgements}
We thank the anonymous referee for comprehensive and constructive feedback that improved this paper.
DWK thanks J. Park for his helpful comments on the use of \texttt{GPCAL}.
This publication is part of the M2FINDERS project which has received funding from the European Research Council (ERC) under the European Union's Horizon 2020 Research and Innovation Programme (grant agreement No 101018682).
This research has made use of data obtained with the Global Millimeter VLBI Array (GMVA), which consists of telescopes operated by the MPIfR, IRAM, Onsala, Mets\"{a}hovi, Yebes, the Korean VLBI Network, the Greenland Telescope, the Green Bank Observatory and the Very Long Baseline Array (VLBA). The VLBA and the GBT are facilities of the National Science Foundation operated under cooperative agreement by Associated Universities, Inc. The data were correlated at the correlator of the MPIfR in Bonn, Germany.
This work is partly based on observations carried out with the IRAM NOEMA Interferometer and the IRAM 30m telescope. IRAM is supported by INSU/CNRS (France), MPG (Germany) and IGN (Spain).
This study makes use of VLBA data from the VLBA-BU Blazar Monitoring Program (BEAM-ME and VLBA-BU-BLAZAR; http://www.bu.edu/blazars/BEAM-ME.html), funded by NASA through the Fermi Guest Investigator Program. The VLBA is an instrument of the National Radio Astronomy Observatory. The National Radio Astronomy Observatory is a facility of the National Science Foundation operated by Associated Universities, Inc.
We reference \citet{abdollahi2023} for use of a $\gamma$-ray light curve presented in the \textsl{Fermi}-LAT Light Curve Repository (LCR). We also refer to the LCR Usage Notes (https://fermi.gsfc.nasa.gov/ssc/data/access/lat/LightCurveRepository/about.html) for important details and caveats about the LCR analysis.
This research was supported by Basic Science Research Program through the National Research Foundation of Korea (NRF) funded by the Ministry of Education (2022R1A6A3A03069095). 
BB acknowledges the financial support of a Otto Hahn research group from the Max Planck Society.

\end{acknowledgements}

%

\begin{thebibliography}{}

\bibitem[Abdollahi et al.(2023)]{abdollahi2023} Abdollahi, S., Ajello, M., Baldini, L., et al. 2023, ApJS, 265, 31

\bibitem[Baczko et al.(2016)]{baczko2016} Baczko, A. -K., Schulz, R., Kadler, M., et al. 2016, A\&A, 593, A47

\bibitem[Blandford et al.(2019)]{blandford2019} Blandford, R., Meier, D., \& Readhead, A. 2019, ARA\&A, 57, 467

\bibitem[Boccardi et al.(2016)]{boccardi2016} Boccardi, B., Krichbaum, T. P., Bach, U., et al. 2016, A\&A, 588, L9

\bibitem[Burd et al.(2022)]{burd2022} Burd, P. R., Kadler, M., Mannheim, K., et al. 2022, A\&A, 660, A1

\bibitem[CASA Team et al.(2022)]{casa2022} CASA Team, Bean, B., Bhatnagar, S., et al. 2022, PASP, 134, 114501

\bibitem[Casadio et al.(2017)]{casadio2017} Casadio, C., Krichbaum, T. P., Marscher, A. P., et al. 2017, Galaxies, 5, 67

\bibitem[Casadio et al.(2021)]{casadio2021} Casadio, C., MacDonald, N. R., Boccardi, B., et al. 2021, A\&A, 649, A153

\bibitem[Cohen et al.(2015)]{cohen2015} Cohen, M. H., Meier, D. L., Arshakian, T. G., et al. 2015, ApJ, 803, 3

\bibitem[Denn et al.(2000)]{denn2000} Denn, G. R., Mutel, R. L., \& Marscher, A. P. 2000, ApJS, 129, 61

\bibitem[G\'{o}mez et al.(2016)]{gomez2016} G\'{o}mez, J. L., Lobanov, A. P., Bruni, G., et al. 2016, ApJ, 817, 96

\bibitem[Greisen(2003)]{greisen2003} Greisen, E. W. 2003, ASSL, 285, 109

\bibitem[Hada et al.(2018)]{hada2018} Hada, K., Doi, A., Wajima, K., et al. 2018, ApJ, 860, 141

\bibitem[Hodgson et al.(2017)]{hodgson2017} Hodgson, J. A., Krichbaum, T. P., Marscher, A. P., et al. 2017, A\&A, 597, A80

\bibitem[Homan et al.(2002)]{homan2002} Homan, D. C., Wardle, J. F. C., Cheung, C. C., et al. 2002, ApJ, 580, 742

\bibitem[Homan et al.(2021)]{homan2021} Homan, D. C., Cohen, M. H., Hovatta, T., et al. 2021, ApJ, 923, 67

\bibitem[H\"{o}gbom(1974)]{hogbom1974} H\"{o}gbom, J. A. 1974, A\&AS, 15, 417

\bibitem[Hovatta et al.(2012)]{hovatta2012} Hovatta, T., Lister, M. L., Aller, M. F., et al. 2012, AJ, 144, 105

\bibitem[Issaoun et al.(2022)]{issaoun2022} Issaoun, S., Wielgus, M., Jorstad, S., et al. 2022, ApJ, 934, 145

\bibitem[Janssen et al.(2019)]{janssen2019} Janssen, M., Goddi, C., van Bemmel, I. M., et al. 2019, A\&A, 626, A75

\bibitem[Janssen et al.(2022)]{janssen2022} Janssen, M., Radcliffe, J. F., Wagner, J. 2022, Universe, 8, 527

\bibitem[Jorstad \& Marscher(2003)]{jorstad2003} Jorstad, S. G. \& Marscher, A. P. 2003, ASPC, 299, 111

\bibitem[Jorstad et al.(2017)]{jorstad2017} Jorstad, S. G., Marscher, A. P., Morozova, D. A., et al. 2017, ApJ, 846, 98

\bibitem[Jorstad et al.(2022)]{jorstad2022} Jorstad, S. G., Marscher, A. P., Raiteri, C. M., et al. 2022, Nature, 609, 265

\bibitem[Jorstad et al.(2023)]{jorstad2023} Jorstad, S. G., Wielgus, M., Lico, R., et al. 2023, ApJ, 943, 170

\bibitem[Kim et al.(2018)]{kim2018} Kim, D. -W., Trippe, S., Lee, S. -S., et al. 2018, MNRAS, 480, 2324

\bibitem[Kim et al.(2019)]{kim2019} Kim, J. -Y., Krichbaum, T. P., Marscher, A. P., et al. 2019, A\&A, 622, A196

\bibitem[Kim et al.(2022)]{kim2022} Kim, D. -W., Kravchenko, E. V., Kutkin, A. M., et al. 2022, ApJ, 925, 64

\bibitem[Kovalev et al.(2020)]{kovalev2020} Kovalev, Y. Y., Pushkarev, A. B., Nokhrina, E. E., et al. 2020, MNRAS, 495, 3576

\bibitem[Krichbaum et al.(1998)]{krichbaum1998} Krichbaum, T. P., Graham, D. A., Witzel, A., et al. 1998, A\&A, 335, L106

\bibitem[Lepp\"{a}nen et al.(1995)]{leppanen1995} Lepp\"{a}nen, K. J., Zensus, J. A., \& Diamond, P. J. 1995, AJ, 110, 2479

\bibitem[Liodakis et al.(2018)]{liodakis2018} Liodakis, I., Hovatta, T., Huppenkothen, D., et al. 2018, ApJ, 866, 137

\bibitem[Lister et al.(2009)]{lister2009} Lister, M. L., Cohen, M. H., Homan, D. C., et al. 2009, AJ, 138, 1874

\bibitem[Lister et al.(2013)]{lister2013} Lister, M. L., Aller, M. F., Aller, H. D., et al. 2013, AJ, 146, 120

\bibitem[Lobanov(1998)]{lobanov1998} Lobanov, A. P. 1998, A\&A, 330, 79

\bibitem[MacDonald et al.(2015)]{macdonald2015} MacDonald, N. R., Marscher, A. P., Jorstad, S. G., et al. 2015, ApJ, 804, 111

\bibitem[Marscher et al.(2008)]{marscher2008} Marscher, A. P., Jorstad, S. G., D`Arcangelo, F. D., et al. 2008, Nature, 452, 966

\bibitem[Marscher(2016)]{marscher2016} Marscher, A. P. 2016, Galax, 4, 37

\bibitem[Mart\'{i}-Vidal et al.(2021)]{mvidal2021} Mart\'{i}-Vidal, I., Mus, A., Janssen, M., et al. 2021, A\&A, 646, A52

\bibitem[Meier et al.(2001)]{meier2001} Meier, D. L., Koide, S., \& Uchida, Y. 2001, Science, 291, 84

\bibitem[Meyer.(2018)]{meyer2018} Meyer, E. T. 2018, NatAs, 2, 32

\bibitem[Mizuno et al.(2011)]{mizuno2011} Mizuno, Y., Hardee, P. E., \& Nishikawa, K. -I. 2011, ApJ, 734, 19

\bibitem[Nair et al.(2019)]{nair2019} Nair, D. G., Lobanov, A. P., Krichbaum, T. P., et al. 2019, A\&A, 622, a92

\bibitem[Nakamura et al.(2010)]{nakamura2010} Nakamura, M., Garofalo, D., \& Meier, D. L. 2010, ApJ, 721, 1783

\bibitem[Paraschos et al.(2022)]{paraschos2022} Paraschos, G. F., Krichbaum, T. P., Kim, J. -Y., et al. 2022, A\&A, 665, A1

\bibitem[Park et al.(2021a)]{park2021a} Park, J., Byun, D. -Y., Asada, K., et al. 2021a, ApJ, 906, 85

\bibitem[Park et al.(2021b)]{park2021b} Park, J., Hada, K., Nakamura, M., et al. 2021b, ApJ, 909, 76

\bibitem[Pushkarev et al.(2009)]{push2009} Pushkarev, A. B., Kovalev, Y. Y., Lister, M. L., et al. 2009, A\&A, 507, L33

\bibitem[Pushkarev et al.(2023)]{push2023} Pushkarev, A. B., Aller, H. D., Aller, M. F., et al. 2023, MNRAS, 520, 6053

\bibitem[Raiteri et al.(2017)]{raiteri2017} Raiteri, C. M., Villata, M., Acosta-Pulido, J. A., et al. 2017, Nature, 552, 374

\bibitem[Rani et al.(2016)]{rani2016} Rani, B., Krichbaum, T. P., Hodgson, J. A., et al. 2016, Galaxies, 4, 32

\bibitem[Readhead(1994)]{rhead1994} Readhead, A. C. S. 1994, ApJ, 426, 51

\bibitem[Ros et al.(2020)]{ros2020} Ros, E., Kadler, M., Perucho, M., et al. 2020, A\&A, 633, L1

\bibitem[Sahakyan et al.(2023)]{sahakyan2023} Sahakyan, N., Harutyunyan, G., \& Israyelyan, D. 2023, MNRAS, 521, 1013

\bibitem[Shepherd et al.(1994)]{shepherd1994} Shepherd, M. C., Pearson, T. J., \& Taylor, G. B. 1994, BAAS, 26, 987

\bibitem[Stirling et al.(2003)]{stirling2003} Stirling, A. M., Cawthorne, T. V., Stevens, J. A., et al. 2003, MNRAS, 341, 405

\bibitem[van Bemmel et al.(2022)]{vbemmel2022} van Bemmel, I. M., Kettenis, M., Des, S., et al. 2022, PASP, 134, 114502

\bibitem[Vermeulen et al.(1995)]{vermeulen1995} Vermeulen, R. C., Ogle, P. M., Tran, H. D., et al. 1995, ApJL, 452, L5

\bibitem[Weaver et al.(2022)]{weaver2022} Weaver, Z. R., Jorstad, S. G., Marscher, A. P., et al. 2022, ApJS, 260, 12

\bibitem[Wehrle et al.(2016)]{wehrle2016} Wehrle, A. E., Grupe, D., Jorstad, S. G., et al. 2016, ApJ, 816, 53

\bibitem[Zensus(1997)]{zensus1997} Zensus, J. A. 1997, ARA\&A, 35, 607

\bibitem[Zhang et al.(2022)]{zhang2022} Zhang, H., Li, X., Giannios, D., et al. 2022, ApJ, 924, 90





\end{thebibliography}
%

\appendix

\section{Data calibration}
\label{sec:calim}
The Radboud PIpeline for the Calibration of high Angular Resolution Data \citep[\texttt{rPICARD};][]{janssen2019} is based on the Common Astronomy Software Applications package \citep[\texttt{CASA};][]{casa2022} with a recent VLBI upgrade \citep{vbemmel2022}.
This is a generic VLBI calibration pipeline with a wide range of input parameters that can be specified for a fine-tune data processing if necessary.
In most cases, only the names of the sources designated as science targets and calibrators have to be specified; \texttt{rPICARD} will then produce fully calibrated data with a single command from the terminal and the user can inspect a wide range of diagnostic plots to flag bad data and change some parameters if necessary (e.g., modeling instrumental effects as being time-dependent instead of being constant over the duration of the observation) for a quick re-run.

The detailed \texttt{rPICARD} steps and assumptions made for the data calibration are given in \citet{janssen2019}. Here, we list the specific procedures employed to calibrate the 24 April 2021 GMVA observation, which is special in the sense that we used BL Lac as both the calibrator source and the science target:
\begin{enumerate}
    \item We used the \texttt{rPICARD} inputs listed under the CDS. We have set a list of refants (i.e., reference antennas), using NOEMA as the primary one and restricted the search range for the optimized solution interval of a segmented fringe-search (for a model-agnostic calibration of the Earth's troposphere) a little bit for computational speedup. We have also raised the default S/N cutoff for fringe-fit of 3.3 to 5.5. The default 3.3 cutoff will yield more detections in the low S/N regime but also a higher probability of false fringes, which would require a more careful inspection and flagging of the data. We note that we already got more detections with a 5.5 cutoff compared to our reference calibration with \texttt{AIPS}, where a 4.5 cutoff was used (see below). Initially, we set no science target
 (``\texttt{None}'') and ``\texttt{BLLAC}'' as a calibrator source.
    \item Instead of running the full pipeline, we ran only the instrumental calibration steps with the \texttt{picard -p -r -q x,0$\sim$10} command.
    \item We then deleted intermediate fringe-fit tables that contained solutions for BL Lac\footnote{In the working directory, \texttt{calibration\_tables/ff\_mb\_cal\_c.t} and \texttt{calibration\_tables/ff\_mb\_cal\_s.t}}. These tables were used for a signal stabilization \citep[correction for atmospheric turbulence,][]{janssen2022} prior to solving for instrumental effects.
    \item Then, we have updated our input files by setting BL Lac as science target and no longer a calibrator source. This is a small change in one of the input files (i.e., a second ``observation.inp''). We have uploaded the updated file to the CDS.
    \item We ran the remaining calibration steps for an optimal calibration of BL Lac, first a wide fringe search over scan durations and then a search with shorter, optimized solution intervals: \texttt{picard -p -r -fm -q x,12$\sim$15}.
\end{enumerate}

\begin{figure}[t]
\centering
\includegraphics[angle=0, width=\columnwidth, keepaspectratio]{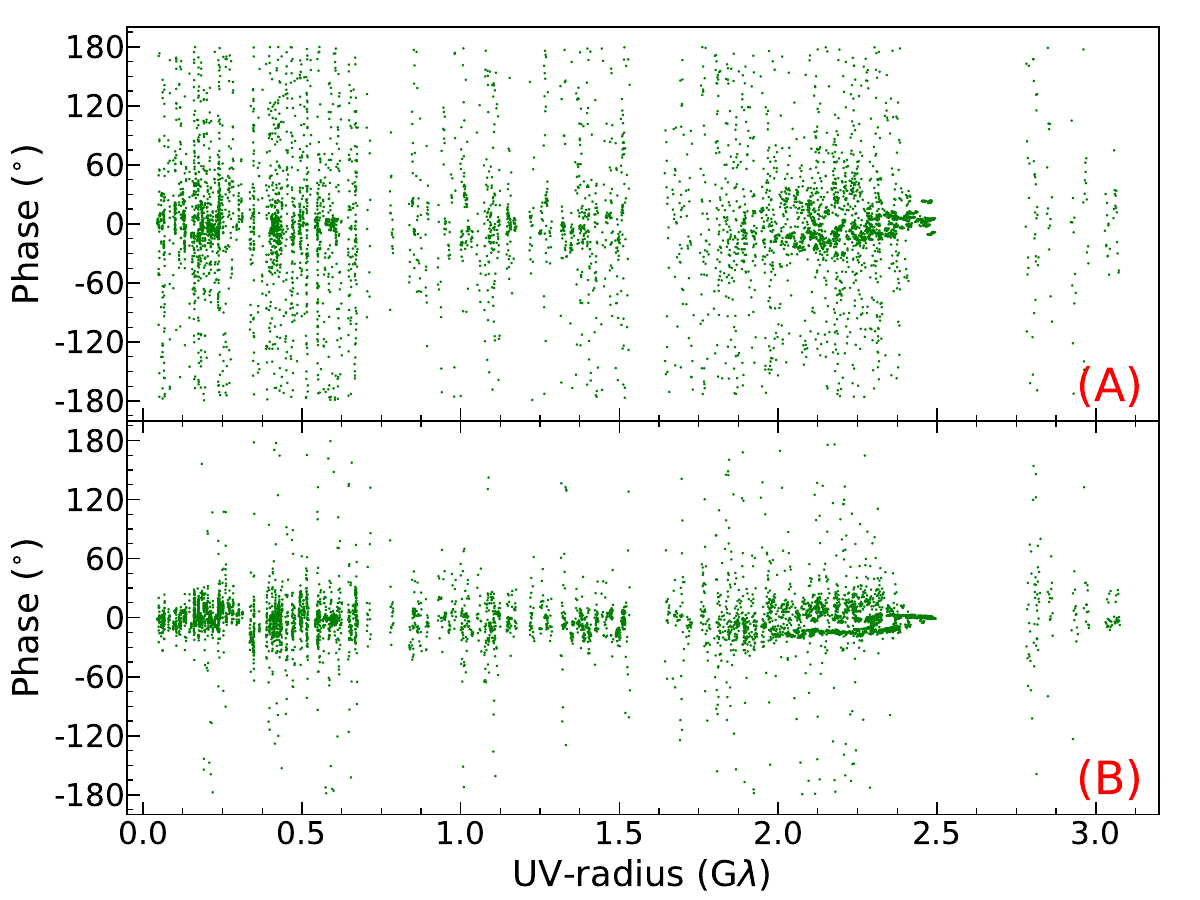}
\caption{Visibility phases of the calibrated data: the conventional manual calibration with \texttt{AIPS} (A) versus the automatic calibration with \texttt{rPICARD} (B); they come before the imaging processes. The data were averaged into 600\,second-bins for clarity.
}
\label{fig:phscomp}
\end{figure}

\begin{figure}[h]
\centering
\includegraphics[angle=0, width=\columnwidth, keepaspectratio]{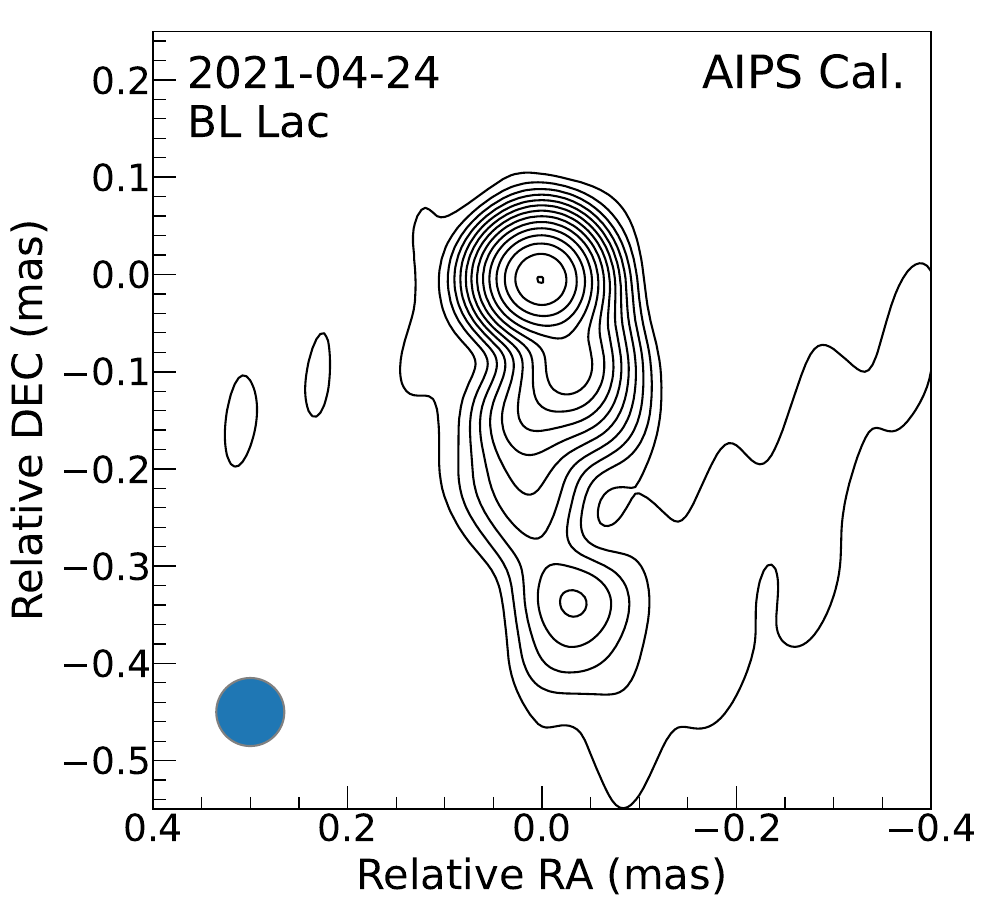}
\caption{
Naturally weighted image of the jet produced by the conventional manual calibration with \texttt{AIPS}.
The map is convolved with a circular restoring beam of radius 0.07\,mas (bottom-left). The contours increase by a factor of $\sqrt{2}$ from 1.1\,\% to 99.6\,\% of the map peak (i.e., $\sim$0.67\,Jy/beam). The rms noise level is around $\sim$3.24\,mJy/beam.
}
\label{fig:aipsim}
\end{figure}

The GMVA datasets have been calibrated in the conventional way of using the Astronomical Image Processing System \citep[\texttt{AIPS};][]{greisen2003} manually. At higher radio frequencies with longer baseline lengths, the coherence time decreases rapidly and one needs more careful, advanced approaches in data calibration. In Figure~\ref{fig:phscomp}, we present visibility phases of our data calibrated by \texttt{rPICARD} and \texttt{AIPS}. For the manual \texttt{AIPS} calibration, we followed a standard procedure for the manual calibration with \texttt{AIPS} \citep[see e.g.,][]{kim2019}: manual phase calibration, global fringe-fitting with a few minutes of the solution interval (four minutes in our case), and amplitude and opacity correction with some metadata (i.e., Tsys measurements, gain curves, and weather information). We carefully inspected the Tsys measurements of all the GMVA antennas and flagged and replaced any erroneous measurements (e.g., 999\,K) with reasonable values by using linear interpolation. 
We found 13968448 and 11634432 visibilities in total from the \texttt{rPICARD} and manual \texttt{AIPS} calibrations, respectively. This means that \texttt{rPICARD} returns $\sim$20\,\% of more visibilities; the S/N cutoff in the fringe-fitting was set to a conservative value of 5.5 for \texttt{rPICARD} and 4.5 for \texttt{AIPS}, thus meaning that the threshold for \texttt{AIPS} was even lower than the case of \texttt{rPICARD}.\footnote{However, we note that there might be small differences in how \texttt{AIPS} and \texttt{CASA} calculate the thermal noise and the fringe-fit FFT S/N.} For clarity, those visibilities in Figure~\ref{fig:phscomp} are averaged with an interval of 600\,seconds. The mean ($\overline{\rm ph}$) and standard deviation (1$\sigma_{\rm ph}$) of the phases are: $\overline{\rm ph}$\,$\sim$\,$-$0.9$^{\circ}$ and 1$\sigma_{\rm ph}$\,$\sim$\,28.5$^{\circ}$ for \texttt{rPICARD} and $\overline{\rm ph}$\,$\sim$\,1.8$^{\circ}$ with 1$\sigma_{\rm ph}$\,$\sim$\,64.1$^{\circ}$ for \texttt{AIPS}. These estimates suggest that \texttt{rPICARD} improved the phase stability by a factor of $>$\,2; for comparison, we also provide an image of the jet generated from the same data, but using the manual \texttt{AIPS} calibration (see Figure~\ref{fig:aipsim}).

We have identified three features of \texttt{rPICARD} that are responsible for the improved data calibration over the standard \texttt{AIPS} procedures typically employed for the calibration of GMVA data:
\begin{enumerate}
    \item The ``coherence'' calibration as part of the signal stabilization prior to solving for instrumental effects substantially reduces the impact of atmospheric phase turbulence and thus leads to more accurate solutions for instrumental calibration steps, such as telescope phase bandpasses.
    \item By combining all correlation products after the instrumental alignments (including RL phases and delays), higher sensitivities for the fringe search are obtained.
    \item For the standard \texttt{AIPS} calibration, a single solution interval is used for the fringe-fitting, namely, four minutes in the case of this data. If that interval is too long, atmospheric phase turbulence will not be corrected well. If it is too short, detections may be lost due to poor S/N. Here, \texttt{rPICARD} uses an adaptive solution interval to achieve an optimal calibration. The segmentation time is tuned for each individual scan over a wide search range (starting with five seconds for this data) by picking the shortest interval that yields the maximum number of detections. In some cases \citep[see][for details]{janssen2019}, different solution intervals can be used for different antennas in the same scan, for instance, when all the data to single antenna has low S/N. The solutions obtained on different intervals are then combined into a single calibration table via interpolation.
\end{enumerate}

\begin{figure}[t]
\centering
\includegraphics[angle=0, width=\columnwidth, keepaspectratio]{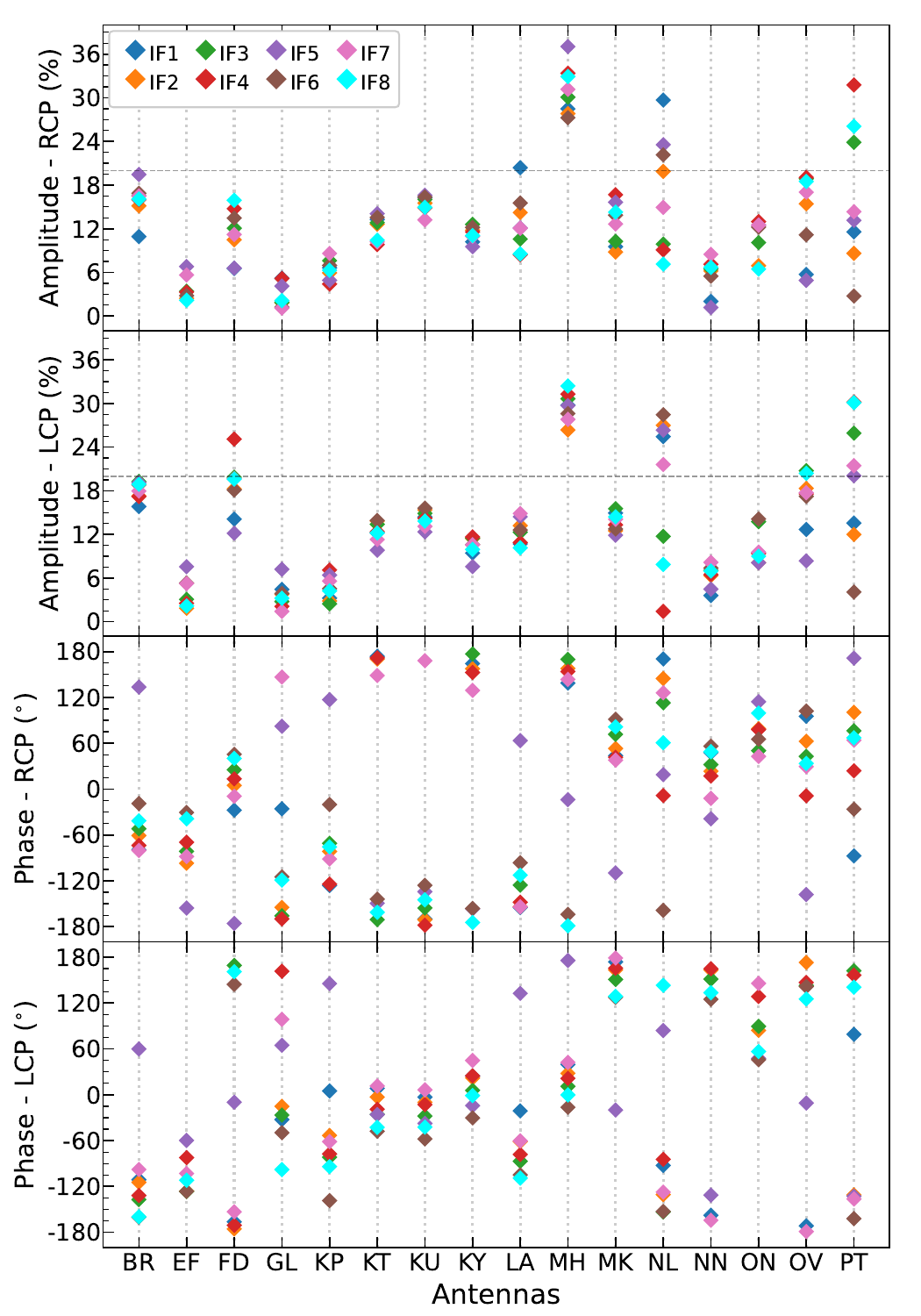}
\caption{D-terms of the GMVA antennas estimated by GPCAL from the data. From top to bottom: RCP/LCP amplitudes and RCP/LCP phases. The horizontal dashed lines in the amplitude panels indicate 20\,\%.
}
\label{fig:dterm}
\end{figure}

Next to a single Measurement Set that contains the calibrated and raw data of all sources, \texttt{rPICARD} generates UVFITS files of the calibrated data for each source in the data after all the calibration steps are finished. Then, this \textsl{uv}-data was imaged with the software \texttt{Difmap} \citep{shepherd1994}. We employed the conventional hybrid imaging approach: cycles of the CLEAN algorithm \citep{hogbom1974} plus amplitude and phase self-calibration, iteratively. To highlight source structures with low-surface brightness and reduce side-lobes, we applied a Gaussian \textsl{uv}-taper to the data at \textsl{uv} radius of 1.2\,G$\lambda$ and the resultant beam size is 0.184\,$\times$\,0.058\,mas at $-$6.09$^{\circ}$. During the imaging processes, we used both the uniform and natural weighting schemes to find any missing fluxes of the jet. In this particular dataset, however, we found that the uniform weighting is better than the natural weighting in terms of the \textsl{rms} noise level: 1$\sigma_{\rm I,rms}$ $\sim$ 2.35\,mJy/beam for natural weighting and 1.29\,mJy/beam for uniform weighting. The overall jet shape appears in the final image more smoothly and clearly with the natural weighting without small-scale artifacts that disturb the edges of the jet. Thus, we employ the naturally weighted image in the analysis of the jet structure.

For polarization calibration (i.e., solving for the D-term), we employed the Generalized Polarization CALibration pipeline \citep[\texttt{GPCAL};][]{park2021a}. \texttt{} It employs both the \texttt{Difmap} and \texttt{AIPS}.
This pipeline was developed recently to overcome a number of limitations of the conventional program \texttt{LPCAL} \citep{leppanen1995} that has been a standard way most widely used in the community for the polarization calibration \citep[see also][for another advanced \texttt{CASA}-based program \texttt{PolSolve}]{mvidal2021}. 
In the calibration, BL\,Lac was used as polarization calibrator. We divided the jet into four separate regions that cover all the CLEAN components in the final image. At such a high frequency (i.e., 86\,GHz), polarization structures of the inner subpc-scale jet are likely to be complicated and the similarity assumption \citep[e.g.,][]{leppanen1995} may not hold. Thus, we employ the instrumental polarization self-calibration technique to reflect those complex polarization structures properly. For more details about \texttt{GPCAL}, we refer to \citet{park2021a}.

Figure~\ref{fig:dterm} shows amplitudes and phases of the D-terms estimated by \texttt{GPCAL}. The following antenna codes are used: VLBA (North Liberty; NL, Fort Davis; FD, Los Alamos; LA, Pie Town; PT, Kitt Peak; KP, Owens Valley; OV, Brewster; BR, and Mauna Kea; MK), KVN (Yonsei; KY, Ulsan; KU, and Tamna; KT), Effelsberg; EF, Onsala; ON, Mets\"{a}hovi; MH, Pico Veleta; PV, NOEMA; NN, and the Greenland Telescope; GL.
For some basic information on the individual antennas, we refer to the "Sensitivities" section in the GMVA website\footnote{https://www3.mpifr-bonn.mpg.de/div/vlbi/globalmm/}.

We found that some IFs and/or antennas are problematic during the imaging process. We lost the first IF (IF1) in EF and ON. Overall, IF5 was less sensitive than the other IFs. As expected from its high Tsys values ($\sim$450\,K on average), MH was very noisy throughout the whole scans and we found mean values of its D-term amplitudes for both RCP and LCP to be $\sim$30\,\%. In each antenna, the D-term amplitudes are quite variable with different IFs, particularly FD, NL, OV, and PT (see Figure~\ref{fig:dterm}). However, the overall fitting processes with \texttt{GPCAL} yielded good results with a mean reduced chi-squared value of $\sim$1.7 for all the IFs. We also find that the D-terms of LCP are slightly higher than the ones of RCP on average. The mean amplitude of the D-terms of RCP over the whole IFs/antennas, is $\sim$12\,\% ($\sim$11\,\% without MH). In the case of LCP, the mean amplitude is $\sim$13\,\% ($\sim$12\,\% without MH). 
We further refer to \citet{casadio2017} for previous D-term estimates of the GMVA antennas. For all the antennas except MH, the overall D-terms are below 20\,\%, with NL \& PT having a number of IFs exceeding 20\,\%; during the data calibration, we noticed that these antennas (including MH) were much worse than the others in this dataset.
The best D-terms were found in EF, GL, KP, and NN that range 3--6\,\% on average. In the polarization imaging, we excluded any IFs or antennas with the D-terms exceeding 20\,\%.
We also note that both with averaged or separated IFs, those polarized features (i.e., P1--4) appeared in the final image. The latter was used in this work due to better \textsl{rms} noise level in polarization.
We further compared this image to the one generated without NL and PT completely and there was no significant difference between them.

Although NOEMA joined the observation, quite a number of the other telescopes were not in good condition, affecting the overall data quality. MH and PV suffered from bad weather; in particular, no fringes were found to PV due to bad weather. 
Unfortunately, most of the VLBA antennas suffered from pointing and/or focus issues in this epoch (plus bad receiver performance in FD and PT).

\section{Gaussian model-fit components}
\label{sec:ccs}

\begin{figure}[t]
\centering
\includegraphics[angle=0, width=\columnwidth, keepaspectratio]{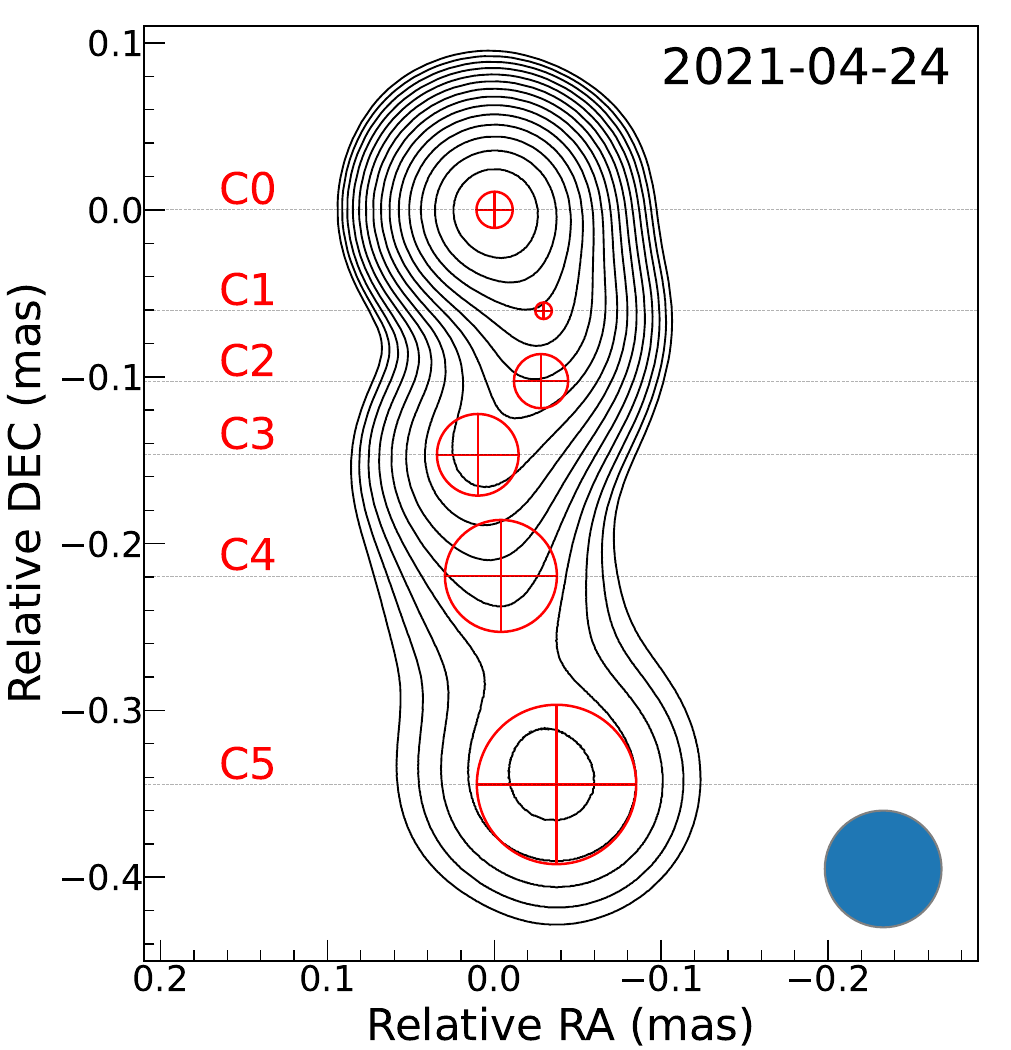}
\caption{Gaussian model-fit image of the jet observed by the GMVA at 86\,GHz. The Gaussian components are shown in red color. The map is convolved with a circular restoring beam of radius 0.07\,mas (bottom-right). The contours increase by a factor of $\sqrt{2}$ from 0.8\,\% to 72.4\,\%.
}
\label{fig:ccs}
\end{figure}

The final image of the jet was also modeled with multiple circular Gaussian components. Figure~\ref{fig:ccs} shows the resultant image of the jet. The model visibilities consist of seven Gaussian components and return a reduced Chi-squared value of $\sim$1.8. All the jet components are labeled by `$C$' plus a number (from 0 to 6). The upstream-end component $C0$ is considered as the radio core. The detection of $C6$ was marginal and is far away from the map center ($\sim$1.6\,mas), compared to the other jet components. This region is not a region of interest in our study, thus excluded in Figure~\ref{fig:ccs}.
The fitted parameters of the components are shown in Table~\ref{tab:ccs}. The uncertainties of the parameters are very small and thus negligible. To be conservative, however, one could consider 10\,\% of the fluxes and $\sim$1/10 of the beam size as the uncertainties for their flux and position, respectively \citep[e.g.,][]{lister2009}. The apparent brightness temperatures of the components are calculated following \citet{issaoun2022}.

\begin{table}[t]
\caption{Properties of the model-fitted Gaussian components.}
\label{tab:ccs}
\centering 
\begin{tabular}{l @{\hspace{4mm}} c @{\hspace{4mm}} c @{\hspace{5mm}} c @{\hspace{7mm}} c @{\hspace{1mm}} c}
\hline      
ID & Flux & Radius\,$^a$ & PA\,$^b$ & FWHM & T$_{\rm b}$\,$^{c}$  \\
  &  (Jy)  &  (mas)  &  ($^{\circ}$)  &  (mas)  &  (10$^{10}$\,K)  \\
\hline
C0  &  0.824  &  0.000  &  -  &  0.022  &  285.6  \\
C1  &  0.169  &  0.067  &  $-$154.0  &  0.010  &  277.8  \\
C2  &  0.067  &  0.106  &  $-$164.8  &  0.033  &  10.3  \\
C3  &  0.096  &  0.147  &  176.1  &  0.049  &  6.5  \\
C4  &  0.038  &  0.219  &  $-$179.0  &  0.067  &  1.4  \\
C5  &  0.078  &  0.346  &  $-$173.8  &  0.096  &  1.4  \\
C6  &  0.019  &  1.575  &  $-$169.6  &  0.071  &  0.6  \\
\hline
\multicolumn{6}{l}{$^a$ Positions of the components with respect to the core.}\\
\multicolumn{6}{l}{$^b$ From North to South: CW (negative) and CCW (positive).}\\
\multicolumn{6}{l}{$^c$ Observed brightness temperature.}\\
\end{tabular}
\end{table}

\section{Analysis of jet transversal structure}
\label{sec:rganal}
The maximum angular resolution of our data is $\sim$40\,$\mu$as. However, the synthesized elliptical beam is highly elongated due to the lack of long \textsl{uv}-spacings in the north-south direction which distort the image largely. We found that a restoring circular beam of 70\,$\mu$as best describes the jet structure.
For both the CLEAN and model-fit images, the map center was aligned to the position of the map peak; the initial position of the peak was southern by $\sim$0.02\,mas from the map center. Using polar coordinates (Clockwise; CW, 0$^{\circ}$ from the left on the image plane), we collected pixel values (i.e., fluxes) along a circle centered at the map center (0, 0). The circle was made with an angle step of 0.5$^{\circ}$, and its radius corresponds to the radial distance ($r$) from the map center. Then, we fitted a single Gaussian form to the collected data (pixel values vs. angles) and found a best-fit position of the Gaussian peak that is considered as a ridge point in this work. For $r\,\geq$\,0.07\,mas, we applied this approach with $r$ increasing by 0.01\,mas at every step. In the inner regions where $r\,<$\,0.7\,mas, we modified the approach slightly due to the limited pixel resolution and the circles with smaller radius, which make the inner ridge line distorted. Hence, we set a new reference point for the center of the circle to be 0.13\,mas from the map center towards radial direction of the first ridge point measured at $r$\,=\,0.07\,mas. Then, the same process used for the outer part was applied with a starting circle with the radius 0.07\,mas, but towards the map center.

We used the model-fit image to measure the jet width ($W$). In the case of the CLEAN map, we found that the Gaussian modeling describes the transverse emission profile poorly in some regions due to some extra features. Plus, the profile of the width measured with the CLEAN map was erratic and a power-law could not fit to it. Thus, the model-fit image was used to find the jet width in this work. The width was estimated assuming the ridge line that we found above to be the central region of the jet. For every ridge point, we drew a line orthogonal to the line between two consecutive ridge points (i.e., at $r$ and $r-0.01$\,mas). Then, we fitted the Gaussian form to the profile of the orthogonal line. The $FWHM$ of the Gaussian was found by 2$\sqrt{2\,\rm ln2}$\,$\times$\,$\sigma_{\rm r}$, where $\sigma_{\rm r}$ being the standard deviation (or the fitted Gaussian radius). We followed \citet{push2009} to calculate the width and the apparent opening angle ($\Theta_{\rm app}$) of the jet: $W$ = $\sqrt{FWHM^{2} - b_{\rm Beam}^{2}}$ and $\Theta_{\rm app}$ = 2\,arctan($W$/2$r$), with $b_{\rm Beam}$ being the restoring beam size.

Figure~\ref{fig:jetpara} shows a number of jet parameters. Overall, we considered the 2$\sigma$ standard deviation of the fit parameters as their uncertainties. The distance between the two ridge lines ($\Delta d_{\rm ridges}$) was calculated using their x,y-positions on the image plane. To evaluate its uncertainty, we employed the error propagation of their positional errors. In the case of $W$, the errorbars were too small. However, we found that the position of the fitted Gaussian peak often has a small offset to the actual position of the brightest pixel in each of the transverse slices. We added this offset to the uncertainties of $W$. The CN ridge was used to find a flux profile along the jet flow. As mentioned in Section~\ref{sec:ridge}, we consider that the CN ridge line describes the jet structure in more detail. 10\% of the flux densities were assumed to be their uncertainties.

\section{Significance of the polarized knots}
\label{sec:polsigtest}

\begin{figure*}[ht!]
\centering
\includegraphics[width=\textwidth]{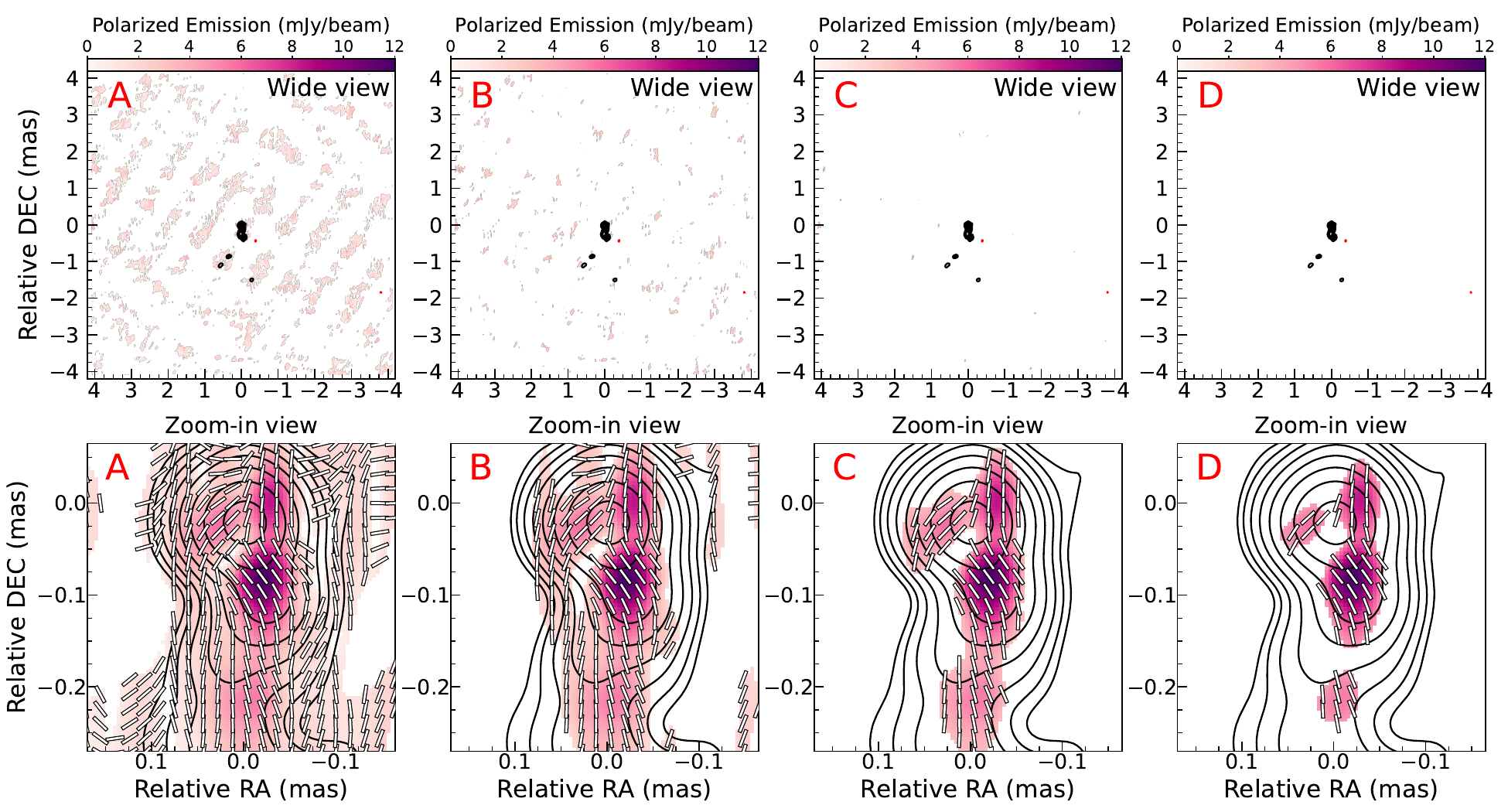}
\caption{Linear polarization maps of the jet with uniform weighting. Different polarization cutoffs are applied to the maps: 1$\sigma_{\rm P,rms}$ for A, 3$\sigma_{\rm P,rms}$ for B, 5$\sigma_{\rm P,rms}$ for C, and 7$\sigma_{\rm P,rms}$ for D. For total intensity, 1$\sigma_{\rm I,rms}$ is used for all the maps. The contour lines represent total intensity increasing by a factor of 2 from 0.6\,\% to 76.8\,\% of the I-peak that is around $\sim$0.8\,Jy. The red contour line represents $-$0.6\,\% of the I-peak. The polarized emission is shown with the color scale. The maps are convolved with a circular restoring beam of radius 0.07\,mas. The white line segments denote EVPAs of the polarized emission.
}
\label{fig:polsig}
\end{figure*}

Depending on the weighting schemes, the residual rms noise level of the polarization map can be estimated as: 1$\sigma_{\rm P,rms}$\,$\sim$\,0.95\,mJy/beam with natural weighting and 0.65\,mJy/beam with uniform weighting. The naturally-weighted image (Figure~\ref{fig:mainpol}) exhibits the four polarized features of the jet on a detection threshold of 4$\sigma_{\rm P,rms}$ for the weakest one (i.e., P4). 
In Figure~\ref{fig:mainpol}, we find the typical values of emission and polarization of the four knots at their central regions (see Table~\ref{tab:polpar}).

In Figure~\ref{fig:polsig}, the uniformly weighted polarization map is examined with different significance levels to filter out any polarized emission weaker than the cutoff (here, P-cut for polarization and I-cut for total intensity).
We set the I-cut to be 1$\sigma_{\rm I,rms}$ for all the maps. At a level of P-cut\,=\,5$\sigma_{\rm P,rms}$, most of the artificial polarized structures have already disappeared from the map and the four polarized knots can be clearly seen. We further checked higher P-cuts. At 7$\sigma_{\rm P,rms}$, only brightest parts of those features are left. Small portions of the P2 and P4 knots still remain in the map up to P-cuts\,=\,8$\sigma_{\rm P,rms}$. For the P1 and P3 knots, they survive until 12$\sigma_{\rm P,rms}$ and 17$\sigma_{\rm P,rms}$, respectively.

\begin{table}
\caption{Flux densities and EVPA of the four polarized knots.}
\label{tab:polpar}
\centering 
\begin{tabular}{c @{\hspace{10mm}} c @{\hspace{10mm}} c @{\hspace{10mm}} c}
\hline      
ID & Stokes-I & Pol. & EVPA\,$^{a}$  \\
  &  (Jy/beam)  &  (mJy/beam)  &  ($^{\circ}$)  \\
\hline
P1  &  0.31  &  6  &  $-$6.5  \\
P2  &  0.50  &  5  &  $-$47.1  \\
P3  &  0.24  &  10  &  36.3  \\
P4  &  0.03  &  5  &  3.1  \\
\hline
\multicolumn{4}{l}{$^a$ From North to South: CW (negative) and CCW (positive).}\\
\end{tabular}
\end{table}

\section{VLBA 86 GHz image in April 2021}
\label{app:bu3mm}

\begin{figure}[b]
\centering
\includegraphics[angle=0, width=\columnwidth, keepaspectratio]{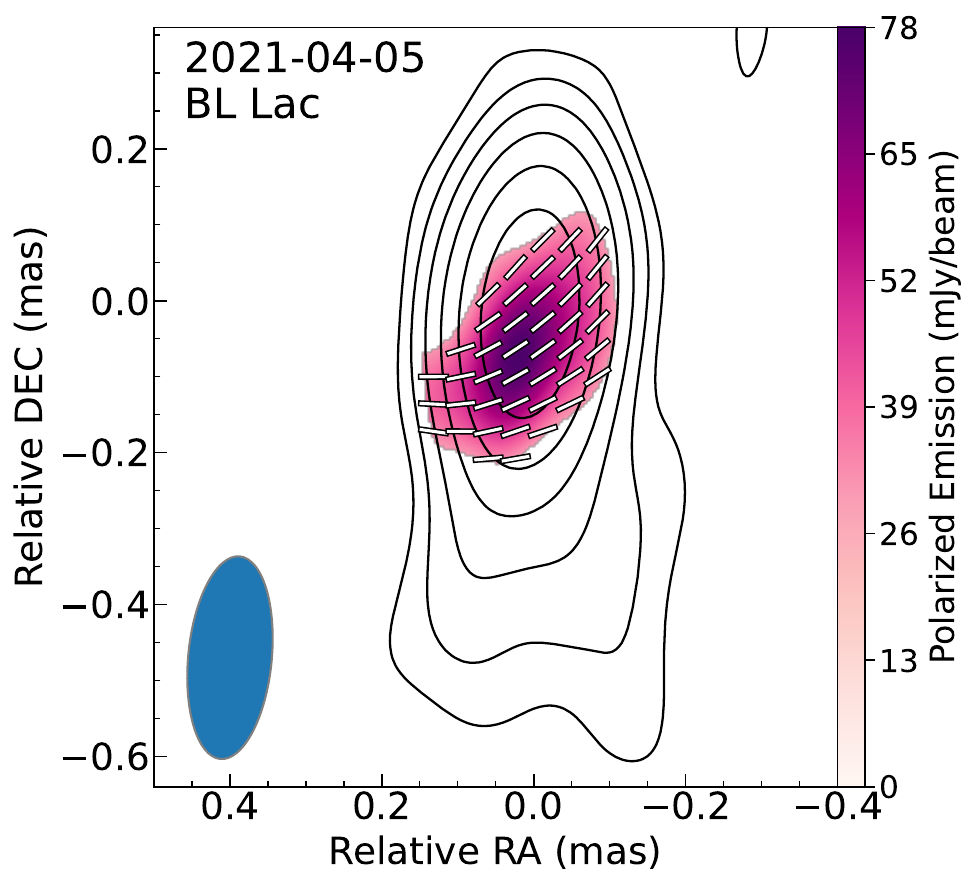}
\caption{BU-VLBA image of the jet at 86\,GHz observed in April 2021. The beam is shown at bottom-left: 0.27\,mas\,$\times$\,0.11\,mas at $-$5.5$^{\circ}$. The contours increase by a factor of 2 from 1.6\,\% to 51.2\,\% of the I-peak that is around $\sim$1.8\,Jy.
}
\label{fig:bu3m}
\end{figure}

BL\,Lac belongs to the samples of the BU monitoring program \citep{jorstad2017}, whose  fully calibrated data (for both 43 \& 86\,GHz) on the BU website\footnote{https://www.bu.edu/blazars/BEAM-ME.html}.
The BU images of the source obtained from near-in-time observations with respect to our GMVA experiment are available.
Figure~\ref{fig:bu3m} shows the BU data of the jet observed on 5 April 2021. The observation was made by the VLBA at 86\,GHz. 
In this VLBA map, we do not find any feature similar to our P3 knot. The central part of the polarized region shows EVPAs almost perpendicular to the jet axis. 
The upper-right and bottom-left edge parts of the extended polarized region seem to be similar in the EVPA orientation to our P1 and P2 knots, respectively. 
Furthermore, we refer to the nearby 7\,mm (43\,GHz) BU images (i.e., two epochs: 5 Apr. \& 28 May in 2021), which hint the presence of P1 and P2 in the upstream region of the jet.
We also note that the EVPAs of our P3 component can be clearly seen in the next-epoch 7\,mm BU image of the jet observed on 28 May 2021. 
It supports that there was a notable change in the core EVPAs (i.e., from perpendicular to parallel to the jet axis) between 5 Apr. 2021 and 24 Apr. 2021.
Hence, we suggest that the jet EVPAs of BL\,Lac at mm-wavelengths can significantly vary within 20\,days near the radio core.

\section{Long-term \texorpdfstring{$\gamma$}{g}-ray light curve of BL\,Lac}
\label{sec:fermi}
\begin{figure}[b]
\centering
\includegraphics[angle=0, width=\columnwidth, keepaspectratio]{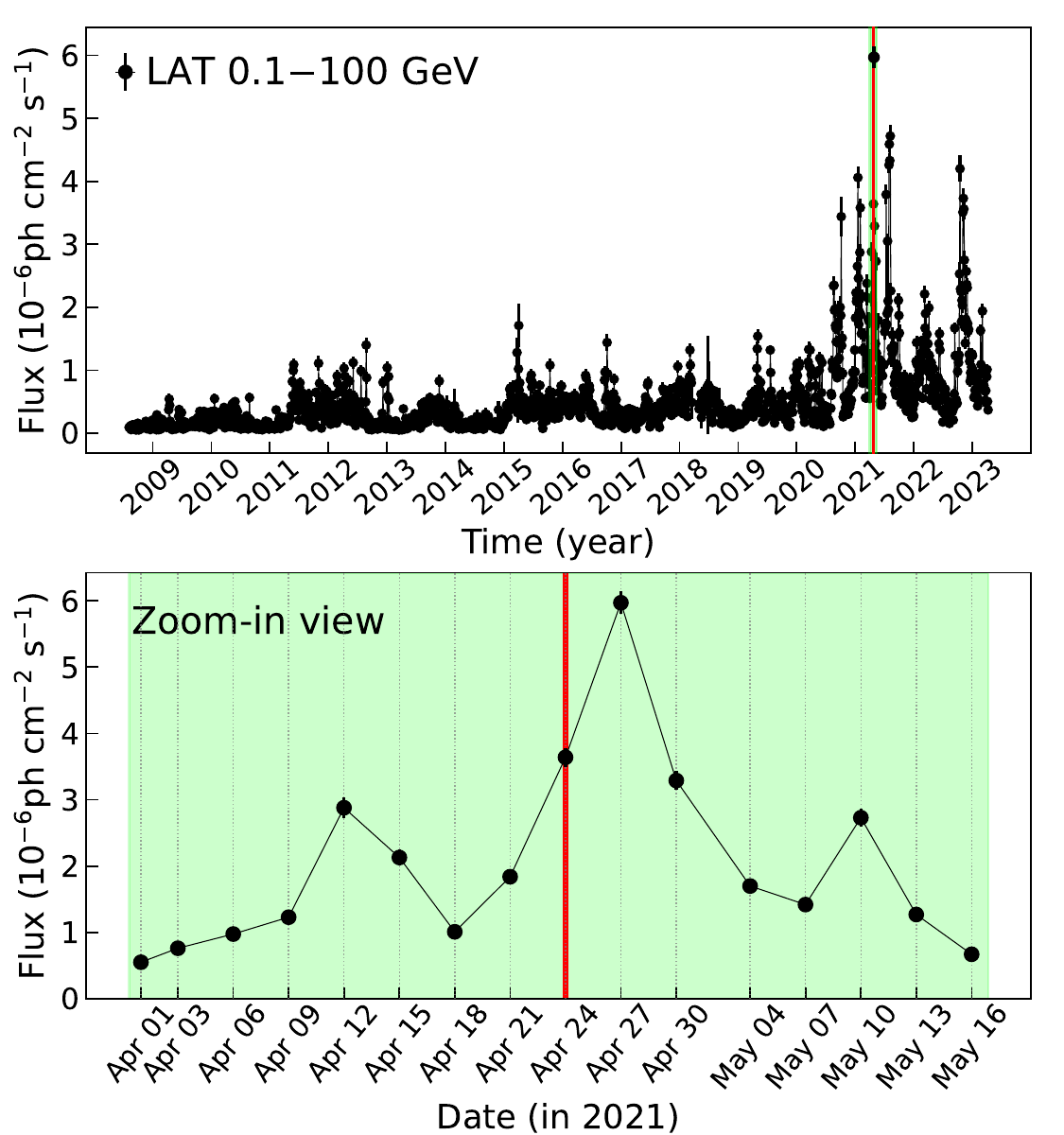}
\caption{
 Long-term $\gamma$-ray light curve of BL\,Lac over 2009--2023 (top).
Zoom-in view of the strongest flare is presented (bottom). 
The red vertical line denotes the observing date of the GMVA observation on 24 April 2021. The green shaded area shows a zoom-in view of the $\gamma$-rays near the GMVA observation.
}
\label{fig:fermi}
\end{figure}

Figure~\ref{fig:fermi} shows a long-term $\gamma$-ray light curve of BL\,Lac obtained from the \textsl{Fermi}-LAT Light Curve Repository \citep[LCR,][]{abdollahi2023}. The light curve spans $\sim$14\,years from 2009 to 2023. The $\gamma$-rays were binned with an interval of 3 days, and only those photons with a test statistic (TS) value above 4 (thus, $\geq$\,2$\sigma$) are included in the light curve; upper limits of the fluxes are excluded from the light curve. For details of the LCR analysis, we refer to the repository webpage\footnote{https://fermi.gsfc.nasa.gov/ssc/data/access/lat/LightCurveRepository/about.html}. The median flux level is 3.5$\rm \,\times\,10^{-7}\,ph\,cm^{-2}\,s^{-1}$ with 1$\sigma$ standard deviation of 1.5$\rm \,\times\,10^{-7}\,ph\,cm^{-2}\,s^{-1}$. In the lower panel of Figure~\ref{fig:fermi}, we show the strongest $\gamma$-ray flare of the source in detail. The flare spans $\sim$10\,days and peaked on 27 April 2021. This peak on 27 April 2021 reaches about $\sim$6.0$\rm \,\times\,10^{-6}\,ph\,cm^{-2}\,s^{-1}$ with TS\,=\,6900, which corresponds to $\sim$83$\sigma$. The GMVA observation then coincides with a rising stage of the flare that is 24 April 2021. This means that our jet image captures a moment of the jet just before the huge $\gamma$-ray outburst. On 24 April 2021, the $\gamma$-ray photon flux density is 
$\sim$3.6$\rm \,\times\,10^{-6}\,ph\,cm^{-2}\,s^{-1}$ with TS\,=\,3200, which corresponds to $\sim$57$\sigma$.

\end{document}